\documentclass[twocolumn]{aa}
    
\usepackage{tikz}
\usetikzlibrary{trees}
\usepackage{listings}
\usepackage{graphicx}
\usepackage{amsmath,amsfonts,amssymb}
\usepackage{txfonts}
\usepackage{color}
\usepackage{natbib}
\usepackage{caption,subcaption}
\usepackage{epsfig}
\usepackage{float}
\usepackage{dblfloatfix}
\usepackage{afterpage}
\usepackage{ifthen}
\usepackage[morefloats=12]{morefloats}
\usepackage{placeins}
\usepackage{multicol}
\usepackage{multirow}
\usepackage{array}
\bibpunct{(}{)}{;}{a}{}{,}
\usepackage[switch]{lineno}
\definecolor{linkcolor}{rgb}{0.6,0,0}
\definecolor{citecolor}{rgb}{0,0,0.75}
\definecolor{urlcolor}{rgb}{0.12,0.46,0.7}
\usepackage[breaklinks, colorlinks, urlcolor=urlcolor,
linkcolor=linkcolor,citecolor=citecolor,pdfencoding=auto]{hyperref}
\hypersetup{linktocpage}

\def\setsymbol#1#2{\expandafter\def\csname #1\endcsname{#2}}
\def\getsymbol#1{\csname #1\endcsname}

\def\Planck{\textit{Planck}}

\newbox\tablebox    \newdimen\tablewidth
\def\leaderfil{\leaders\hbox to 5pt{\hss.\hss}\hfil}
\def\tablenote#1 #2\par{\begingroup \parindent=0.8em
    \abovedisplayshortskip=0pt\belowdisplayshortskip=0pt
    \noindent
    $$\hss\vbox{\hsize\tablewidth \hangindent=\parindent \hangafter=1 \noindent
    \hbox to \parindent{$^#1$\hss}\strut#2\strut\par}\hss$$
    \endgroup}

\def\L2{\ifmmode L_2\else $L_2$\fi}

\def\DeltaT{\ifmmode \Delta T\else $\Delta T$\fi}
\def\deltat{\ifmmode \Delta t\else $\Delta t$\fi}
\def\fknee{\ifmmode f_{\rm knee}\else $f_{\rm knee}$\fi}
\def\Fmax{\ifmmode F_{\rm max}\else $F_{\rm max}$\fi}
\def\solar{\ifmmode{\rm M}_{\mathord\odot}\else${\rm M}_{\mathord\odot}$\fi}
\def\Msolar{\ifmmode{\rm M}_{\mathord\odot}\else${\rm M}_{\mathord\odot}$\fi}
\def\Lsolar{\ifmmode{\rm L}_{\mathord\odot}\else${\rm L}_{\mathord\odot}$\fi}
\def\inv{\ifmmode^{-1}\else$^{-1}$\fi}
\def\mo{\ifmmode^{-1}\else$^{-1}$\fi}
\def\sup#1{\ifmmode ^{\rm #1}\else $^{\rm #1}$\fi}
\def\expo#1{\ifmmode \times 10^{#1}\else $\times 10^{#1}$\fi}
\def\,{\thinspace}
\def\lsim{\mathrel{\raise .4ex\hbox{\rlap{$<$}\lower 1.2ex\hbox{$\sim$}}}}
\def\gsim{\mathrel{\raise .4ex\hbox{\rlap{$>$}\lower 1.2ex\hbox{$\sim$}}}}

\def\simprop{\mathrel{\raise .4ex\hbox{\rlap{$\propto$}\lower 1.2ex\hbox{$\sim$}}}}
\def\deg{\ifmmode^\circ\else$^\circ$\fi}
\def\pdeg{\ifmmode $\setbox0=\hbox{$^{\circ}$}\rlap{\hskip.11\wd0 .}$^{\circ}
          \else \setbox0=\hbox{$^{\circ}$}\rlap{\hskip.11\wd0 .}$^{\circ}$\fi}
\def\arcs{\ifmmode {^{\scriptstyle\prime\prime}}
          \else $^{\scriptstyle\prime\prime}$\fi}
\def\arcm{\ifmmode {^{\scriptstyle\prime}}
          \else $^{\scriptstyle\prime}$\fi}
\newdimen\sa  \newdimen\sb
\def\parcs{\sa=.07em \sb=.03em
     \ifmmode \hbox{\rlap{.}}^{\scriptstyle\prime\kern -\sb\prime}\hbox{\kern -\sa}
     \else \rlap{.}$^{\scriptstyle\prime\kern -\sb\prime}$\kern -\sa\fi}
\def\parcm{\sa=.08em \sb=.03em
     \ifmmode \hbox{\rlap{.}\kern\sa}^{\scriptstyle\prime}\hbox{\kern-\sb}
     \else \rlap{.}\kern\sa$^{\scriptstyle\prime}$\kern-\sb\fi}
\def\ra[#1 #2 #3.#4]{#1\sup{h}#2\sup{m}#3\sup{s}\llap.#4}
\def\dec[#1 #2 #3.#4]{#1\deg#2\arcm#3\arcs\llap.#4}
\def\deco[#1 #2 #3]{#1\deg#2\arcm#3\arcs}
\def\rra[#1 #2]{#1\sup{h}#2\sup{m}}

\def\dots{\relax\ifmmode \ldots\else $\ldots$\fi}
\def\WHzsr{\ifmmode $W\,Hz\mo\,sr\mo$\else W\,Hz\mo\,sr\mo\fi}
\def\mHz{\ifmmode $\,mHz$\else \,mHz\fi}
\def\GHz{\ifmmode $\,GHz$\else \,GHz\fi}
\def\mKs{\ifmmode $\,mK\,s$^{1/2}\else \,mK\,s$^{1/2}$\fi}
\def\muKs{\ifmmode \,\mu$K\,s$^{1/2}\else \,$\mu$K\,s$^{1/2}$\fi}
\def\muKRJs{\ifmmode \,\mu$K$_{\rm RJ}$\,s$^{1/2}\else \,$\mu$K$_{\rm RJ}$\,s$^{1/2}$\fi}
\def\muKHz{\ifmmode \,\mu$K\,Hz$^{-1/2}\else \,$\mu$K\,Hz$^{-1/2}$\fi}
\def\MJysr{\ifmmode \,$MJy\,sr\mo$\else \,MJy\,sr\mo\fi}
\def\MJysrmK{\ifmmode \,$MJy\,sr\mo$\,mK$_{\rm CMB}\mo\else \,MJy\,sr\mo\,mK$_{\rm CMB}\mo$\fi}
\def\microns{\ifmmode \,\mu$m$\else \,$\mu$m\fi}

\def\muK{\ifmmode \,\mu$K$\else \,$\mu$\hbox{K}\fi}
\def\microK{\ifmmode \,\mu$K$\else \,$\mu$\hbox{K}\fi}
\def\muW{\ifmmode \,\mu$W$\else \,$\mu$\hbox{W}\fi}
\def\kms{\ifmmode $\,km\,s$^{-1}\else \,km\,s$^{-1}$\fi}
\def\kmsMpc{\ifmmode $\,\kms\,Mpc\mo$\else \,\kms\,Mpc\mo\fi}
\providecommand{\sorthelp}[1]{}

\def\WMAP{\emph{WMAP}}

\def\LCDM{$\Lambda$CDM}

\def\commander{\texttt{Commander}}

\def\commanderthree{\texttt{Commander3}}

\makeatletter
\ifcase \@ptsize \relax
\newcommand{\miniscule}{\@setfontsize\miniscule{5}{6}}
\or
\newcommand{\miniscule}{\@setfontsize\miniscule{6}{7}}
\or
\newcommand{\miniscule}{\@setfontsize\miniscule{6}{7}}
\fi
\makeatother

\renewcommand{\d}[0]{\vec{d}}

\newcommand{\n}[0]{\vec{n}}

\definecolor{orange}{RGB}{255,127,0}

\newcommand{\s}[0]{\vec{s}}
\renewcommand{\a}[0]{\vec{a}}

\newcommand{\B}[0]{\tens{B}}

\renewcommand{\L}[0]{\tens{L}}
\newcommand{\g}[0]{\vec{g}}

\newcommand{\N}[0]{\tens{N}}

\newcommand{\M}[0]{\tens{M}}

\renewcommand{\P}[0]{\tens{P}}

\newcommand{\Dbp}[0]{\Delta_{\mathrm{bp}}}

\newcommand{\BP}{\textsc{BeyondPlanck}}

\def\inv{^{-1}}

\begin{document}
	
\title{\BP\ IV. On end-to-end simulations in CMB analysis ---\\ Bayesian versus frequentist statistics}
\newcommand{\oslo}[0]{1}
\newcommand{\usp}[0]{2}
\newcommand{\ubc}[0]{3}
\newcommand{\tohoku}[0]{4}
\newcommand{\iia}[0]{5}
\newcommand{\ucb}[0]{6}
\newcommand{\inpe}[0]{7}
\newcommand{\ucl}[0]{8}
\newcommand{\rhodes}[0]{9}
\newcommand{\planetek}[0]{10}
\newcommand{\milanoA}[0]{11}
\newcommand{\milanoB}[0]{12}
\newcommand{\milanoC}[0]{13}
\newcommand{\triesteA}[0]{14}
\newcommand{\princeton}[0]{15}
\newcommand{\ipmu}[0]{16}
\newcommand{\jpl}[0]{17}
\newcommand{\helsinkiA}[0]{18}
\newcommand{\helsinkiB}[0]{19}
\newcommand{\nersc}[0]{20}
\newcommand{\haverford}[0]{21}
\newcommand{\mpa}[0]{22}
\newcommand{\triesteB}[0]{23}
\author{\small
M.~Brilenkov\inst{\oslo}\thanks{Corresponding author: M.~Brilenkov; \url{maksym.brilenkov@astro.uio.no}}
\and
\textcolor{black}{K.~S.~F.~Fornazier}\inst{\usp}
\and
L.~T.~Hergt\inst{\ubc}
\and
G.~A.~Hoerning\inst{\usp}
\and
A. Marins\inst{\usp}
\and
T.~Murokoshi\inst{\tohoku}
\and
F.~Rahman\inst{\iia}
\and
N.-O.~Stutzer\inst{\oslo}
\and
Y. Zhou\inst{\ucb}
\and
F.~.B.~Abdalla\inst{\usp, \inpe, \ucl, \rhodes}
\and
K.~J.~Andersen\inst{\oslo}
\and
\textcolor{black}{R.~Aurlien}\inst{\oslo}
\and
\textcolor{black}{R.~Banerji}\inst{\oslo}
\and
A.~Basyrov\inst{\oslo}
\and
A.~Battista\inst{\planetek}
\and
M.~Bersanelli\inst{\milanoA, \milanoB, \milanoC}
\and
S.~Bertocco\inst{\triesteB}
\and
S.~Bollanos\inst{\planetek}
\and
L.~P.~L.~Colombo\inst{\milanoA,\milanoC}
\and
H.~K.~Eriksen\inst{\oslo}
\and
J.~R.~Eskilt\inst{\oslo}
\and
\textcolor{black}{M.~K.~Foss}\inst{\oslo}
\and
C.~Franceschet\inst{\milanoA,\milanoC}
\and
\textcolor{black}{U.~Fuskeland}\inst{\oslo}
\and
S.~Galeotta\inst{\triesteB}
\and
M.~Galloway\inst{\oslo}
\and
S.~Gerakakis\inst{\planetek}
\and
E.~Gjerl{\o}w\inst{\oslo}
\and
\textcolor{black}{B.~Hensley}\inst{\princeton}
\and
\textcolor{black}{D.~Herman}\inst{\oslo}
\and
\textcolor{black}{T.~D.~Hoang}\inst{\ipmu}
\and
M.~Ieronymaki\inst{\planetek}
\and
\textcolor{black}{H.~T.~Ihle}\inst{\oslo}
\and
J.~B.~Jewell\inst{\jpl}
\and
\textcolor{black}{A.~Karakci}\inst{\oslo}
\and
E.~Keih\"{a}nen\inst{\helsinkiA, \helsinkiB}
\and
R.~Keskitalo\inst{\nersc}
\and
G.~Maggio\inst{\triesteB}
\and
D.~Maino\inst{\milanoA, \milanoB, \milanoC}
\and
M.~Maris\inst{\triesteB}
\and
S.~Paradiso\inst{\milanoA}
\and
B.~Partridge\inst{\haverford}
\and
M.~Reinecke\inst{\mpa}
\and
A.-S.~Suur-Uski\inst{\helsinkiA, \helsinkiB}
\and
T.~L.~Svalheim\inst{\oslo}
\and
D.~Tavagnacco\inst{\triesteB, \triesteA}
\and
H.~Thommesen\inst{\oslo}
\and
M.~Tomasi\inst{\milanoA, \milanoB}
\and
D.~J.~Watts\inst{\oslo}
\and
I.~K.~Wehus\inst{\oslo}
\and
A.~Zacchei\inst{\triesteB}
}
\institute{\scriptsize
Institute of Theoretical Astrophysics, University of Oslo, Blindern, Oslo, Norway\goodbreak
\and
Instituto de F\'isica, Universidade de S\~ao Paulo - C.P. 66318, CEP: 05315-970, S\~ao Paulo, Brazil\goodbreak
\and
Department of Physics and Astronomy, University of British Columbia, Vancouver, BC~V6T\,1Z1, Canada\goodbreak
\and
Astronomical Institute, Tohoku University, Sendai, Miyagi 980-8578, Japan\goodbreak
\and
Indian Institute of Astrophysics, Koramangala II Block, Bangalore, 560034, India\goodbreak
\and
Department of Physics, University of California, Berkeley, Berkeley, California, U.S.A.\goodbreak
\and
Instituto Nacional de Pesquisas Espaciais, Divisão de Astrofísica, Av. dos Astronautas, 1758, 12227-010 - São José dos Campos, SP, Brazil\goodbreak
\and
Department of Physics and Astronomy, University College London, Gower Street, London, WC1E 6BT, UK\goodbreak
\and
Department of Physics and Electronics, Rhodes University, PO Box 94, Grahamstown, 6140, South Africa\goodbreak
\and
Planetek Hellas, Leoforos Kifisias 44, Marousi 151 25, Greece\goodbreak
\and
Dipartimento di Fisica, Universit\`{a} degli Studi di Milano, Via Celoria, 16, Milano, Italy\goodbreak
\and
INAF/IASF Milano, Via E. Bassini 15, Milano, Italy\goodbreak
\and
INFN, Sezione di Milano, Via Celoria 16, Milano, Italy\goodbreak
\and
INAF - Osservatorio Astronomico di Trieste, Via G.B. Tiepolo 11, Trieste, Italy\goodbreak
\and
Department of Astrophysical Sciences, Princeton University, Princeton, NJ 08544,
U.S.A.\goodbreak
\and
Kavli IPMU (WPI), UTIAS, The University of Tokyo, Kashiwa, Chiba 277-8583, Japan\goodbreak
\and
 Jet Propulsion Laboratory, California Institute of Technology, 4800 Oak Grove Drive, Pasadena, California, U.S.A.\goodbreak
\and
Department of Physics, Gustaf H\"{a}llstr\"{o}min katu 2, University of Helsinki, Helsinki, Finland\goodbreak
\and
Helsinki Institute of Physics, Gustaf H\"{a}llstr\"{o}min katu 2, University of Helsinki, Helsinki, Finland\goodbreak
\and
Computational Cosmology Center, Lawrence Berkeley National Laboratory, Berkeley, California, U.S.A.\goodbreak
\and
Haverford College Astronomy Department, 370 Lancaster Avenue, Haverford, Pennsylvania, U.S.A.\goodbreak
\and
Max-Planck-Institut f\"{u}r Astrophysik, Karl-Schwarzschild-Str. 1, 85741 Garching, Germany\goodbreak
\and
Dipartimento di Fisica, Universit\`{a} degli Studi di Trieste, via A. Valerio 2, Trieste, Italy\goodbreak
}

\authorrunning{Planck Collaboration}
\titlerunning{Bayesian versus frequentist simulations}

\abstract{End-to-end simulations play a key role in the analysis of
  any high-sensitivity CMB experiment, providing high-fidelity
  systematic error propagation capabilities unmatched by any other
  means. In this paper, we address an important issue regarding such
  simulations, namely how to define the inputs in terms of sky model
  and instrument parameters. These may either be taken as a
  constrained realization derived from the data, or as a random
  realization independent from the data. We refer to these as Bayesian
  and frequentist simulations, respectively. We show that the two
  options lead to significantly different correlation structures, as
  frequentist simulations, contrary to Bayesian simulations,
  effectively include cosmic variance, but exclude
  realization-specific correlations from non-linear
  degeneracies. Consequently, they quantify fundamentally different
  types of uncertainties, and we argue that they therefore also have
  different and complementary scientific uses, even if this dichotomy
  is not absolute; Bayesian simulations are in general more convenient
  for parameter estimation studies, while frequentist simulations are
  in general more convenient for model testing. Before \BP, most
  pipelines have used a mix of constrained and random inputs, and used
  the same hybrid simulations for all applications, even though the
  statistical justification for this is not always
  evident. \BP\ represents the first end-to-end CMB simulation
  framework that is able to generate both types of simulations, and
  these new capabilities have brought this topic to the forefront. The
  Bayesian \BP\ simulations and their uses are described extensively
  in a suite of companion papers. In this paper we consider one
  important applications of the corresponding frequentist simulations,
  namely code validation. That is, we generate a set of 1-year LFI
  30\,GHz frequentist simulations with known inputs, and use these to
  validate the core low-level \BP\ algorithms; gain estimation,
  correlated noise estimation, and mapmaking.}


\maketitle

\tableofcontents
	


\section{Introduction}
\label{sec:introduction}
	
High-fidelity end-to-end simulations play a critical role in the
analysis of any modern CMB experiment for at least three important
reasons. Firstly, during the design phase of the experiment,
simulations are used to optimize and forecast the performance of a
given experimental design, and ensure that the future experiment will
achieve its scientific goals \citep[e.g.,][]{ptep}. Secondly,
simulations are essential for validation purposes, as they may be used
to test data-processing techniques as applied to a realistic
instrument model. Thirdly, realistic end-to-end simulations play an
important role in bias and error estimation for traditional CMB
analysis pipelines.

Simulations played a particularly important role in the data reduction
of \Planck, and massive efforts were invested in implementing
efficient and re-usable analysis codes that were generally applicable
to a wide range of experiments. This work started with the LevelS software package \citep{levelS} and culminated with the \emph{Time Ordered Astrophysics Scalable Tools}\footnote{\url{https://github.com/hpc4cmb/toast}}
(TOAST), which was
explicitly designed to operate in a massively parallel
high-performance computing environment. TOAST was used to produce the
final generations of the \Planck\ \emph{Full Focal Plane} (FFP)
simulations \citep{planck2014-a14}, which served as the main error
propagation mechanism in the \Planck\ 2015 and 2018 data releases
\citep{planck2014-a01,planck2016-l01}.

For \Planck, generating end-to-end simulations represented by far the
dominant computational cost of the entire experiment, accounting for
25 million CPU in the 2015 data release alone. In addition,
the production phase required massive amounts of human effort, in
terms of preparing the inputs, executing the runs, and validating the
outputs. It is of great interest for any future experiment
to optimize and streamline this simulation process, and reuse both
validated software and human work whenever possible.

In this respect, the \BP\ end-to-end Bayesian analysis framework
\citep{bp01} offers a novel approach to generating CMB
simulations. While the primary goal of this framework is to draw
samples from a full joint posterior distribution for analysis
purposes, it is useful to note that the foundation of this approach is
simply a general and explicit parametric model for the full
time-ordered data (TOD). When exploring the full joint posterior
distribution, this model is compared with the observed data in TOD
space. The analysis phase is as such numerically equivalent to
producing a large number of TOD simulations, and comparing each of
these with the actual observed data. In this framework, each step of
the analysis and simulation pipelines are thus fully equivalent, and
the primary difference is simply whether the input model parameters
are assumed to be constrained by the data or not.

This latter observation is in fact a key point regarding end-to-end
simulations for CMB experiments in general, and a main goal of the
current paper is to clarify the importance of choosing input
parameters for a given simulation appropriately. Specifically, we
argue in this paper that two fundamentally different choices are
available; one can either choose parameters that are constrained
directly by the observed data (as is traditionally done for the CMB
Solar dipole or astrophysical foregrounds), or one can choose
parameters that are independent from the observed data (as is
traditionally done for CMB fluctuations or instrumental noise). We
further argue that this choice will have direct consequences for what
scientific questions the resulting simulations are optimized to
address.

It is important to note that these ideas were discussed broadly, but
not systematically, within the \Planck\ community before building the
FFP simulations. For instance, one proposal was to base the
large-scale CMB temperature fluctuations at $\ell\le70$ from
constrained \WMAP\ realizations \citep{bennett2012}, and thereby
integrate knowledge about the real sky into the simulations. Another
proposal was to use the actually observed LFI gain measurements to
generate the simulations. A third and long-standing discussion
revolved around which values to adopt for the CMB Solar dipole.

The \BP\ framework offers a novel systematic view on these questions,
as our Bayesian approach provides for the first time
statistically well-defined constrained realizations for \emph{all}
parameters in the sky model, and not only a small subset. Furthermore,
when comparing the correlation structures that arise from the
posterior samples with those derived from traditional simulations,
obvious and important differences appear, both in terms of frequency
maps \citep{bp10} and CMB maps \citep{bp11}. 

The first main goal of
the current paper is to explain these differences intuitively, and in
that process we introduce the concepts of ``Bayesian simulations'' and
``frequentist simulations''. Bayesian simulations are identical to the
posterior samples described by \citet{bp01}, and represent simulations
that are constrained by the observed data. In contrast, frequentist
simulations are unconstrained by the data.

The second main goal of this paper is simply to demonstrate in
practice how the \BP\ machinery may be used to generate frequentist
simulations, on a similar footing as TOAST, and we will use these
simulations for one important application, namely code validation; as
discussed by \citet{bp03} and \citet{bp05}, the \commander\ code that
forms the computational basis of the \BP\ pipeline is explicitly
designed to be re-used for a wide range of experiments. It is
therefore critically important that this implementation is thoroughly
validated with respect to statistical bias and uncertainties, and we
do that by analyzing well-controlled simulations in this paper.


The rest of the paper is organized as follows. We first provide a
brief overview of the \BP\ framework and data model in
Sect.~\ref{sec:pipeline}. In Sect.~\ref{sec:frequentist}, we introduce
the concept of Bayesian and frequentist simulations, and we discuss
their difference. In Sect.~\ref{sec:simulation} we describe the input
parameters and simulation configuration used in this paper, before
using these simulations to validate the \BP\ implementation in
Sect.~\ref{sec:validation}.
We conclude in Sect.~\ref{sec:conclusions}.


\section{\BP\ data model and Gibbs sampler}
\label{sec:pipeline}

As described in \citet{bp01} and its companion papers, the single most
fundamental component of the \BP\ framework is an explicit parametric
model that is to be fitted to raw TOD that includes instrumental,
astrophysical, and cosmological parameters. For the current analysis,
this model takes the following form,
{\small
\begin{equation}
  \begin{split}
    d_{j,t} = g_{j,t}&\P_{tp,j}\left[\B^{\mathrm{symm}}_{pp',j}\sum_{c}
      \M_{cj}(\beta_{p'}, \Dbp^{j})a^c_{p'}  + \B^{4\pi}_{j,t}\s^{\mathrm{orb}}_{j}  
      + \B^{\mathrm{asymm}}_{j,t} \s^{\mathrm{fsl}}_{t} \right]
    +   \label{eq:todmodel} \\
    + &a_{\mathrm{1Hz}}\s^{\mathrm{1Hz}}_{j} + n^{\mathrm{corr}}_{j,t} + n^{\mathrm{w}}_{j,t}.
  \end{split}
\end{equation}
}
sample, $p$ denotes a single pixel on the sky, and $c$ represents one
single astrophysical signal component. Furthermore, $d_{j,t}$ denotes
the measured data; $g_{j,t}$ denotes the instrumental gain;
$\P_{tp,j}$ is a pointing matrix; $\B_{pp',j}$ denotes beam
convolution with either the (symmetric) main beam, the (asymmetric)
far sidelobes, or the full $4\pi$ beam response; $\M_{cj}(\beta_{p},
\Dbp)$ denotes the so-called mixing matrix, which describes the
amplitude of component $c$ as seen by radiometer $j$ relative to some
reference frequency when assuming some set of bandpass
correction parameters $\Dbp$; $a^c_{p}$ is the amplitude of component
$c$ in pixel $p$; $s^{\mathrm{orb}}_{j,t}$ is the orbital CMB dipole
signal, including relativistic quadrupole corrections;
$s^{\mathrm{fsl}}_{j,t}$ denotes the contribution from far sidelobes;
$s^{\mathrm{1Hz}}_{j,t}$ denotes the contribution from electronic
1\,Hz spikes; $n^{\mathrm{corr}}_{j,t}$ denotes correlated
instrumental noise; and $n^{\mathrm{w}}_{j,t}$ is uncorrelated (white)
instrumental noise. The sky model, denoted by the sum over components,
$c$, in the above expression may be written out as an explicit sum over CMB, synchrotron, free-free, AME, thermal dust, and point source emission, as described by \citet{bp13,bp14}.

On the instrumental side, the correlated noise is associated with a
covariance matrix, $\N^{\mathrm{corr}} = \left<\n^{\mathrm{corr}}
(\n^{\mathrm{corr}})^T \right>$, which may be approximated as piecewise
stationary, and with a Fourier space power spectral density (PSD),
$\N_{ff'} = P(f)\delta_{ff'}$, that for \BP\ consists of a sum of a classic
$1/f$ term and a log-normal term \citep{bp06}, {\small
\begin{equation}
        P(f) = \sigma_0^2\left[1 +
        \left(\frac{f}{f_\mathrm{knee}}\right)^\alpha\right] + A_\mathrm{p} \exp\left[-\frac{1}{2}\left(\frac{\log_{10}f - \log_{10} f_\mathrm{p}}{\sigma_\mathrm{dex}}\right)^2\right].
        \label{eq:1fmodel_lognorm}
\end{equation}}
We define $\xi_n=\{\sigma_0,\alpha,f_{\mathrm{knee}},A_p\}$ as a
composite parameter that is internally sampled iteratively through an
individual Gibbs step, as described by \citet{bp06}; the peak location
and width parameters of the log-normal term, $f_{\mathrm{p}}$ and
$\sigma_\mathrm{dex}$, are currently fixed at representative values.

Denoting the set of all free parameters in
Eqs.~\eqref{eq:todmodel}--\eqref{eq:1fmodel_lognorm} by $\omega$, we
can simplify Eq.~\eqref{eq:todmodel} symbolically to
\begin{equation}
  d_{j,t} = s^{\mathrm{tot}}_{j,t}(\omega) + n^{\mathrm{w}}_{j,t}.
  \label{eq:simplemodel}
\end{equation}
The \BP\ approach to CMB analysis simply amounts to mapping out the
posterior distribution as given by Bayes' theorem,
\begin{equation}
  P(\omega\mid \d) = \frac{P(\d\mid \omega)P(\omega)}{P(\d)} \propto
  \mathcal{L}(\omega)P(\omega),
  \label{eq:jointpost}
\end{equation}
where $P(\d\mid \omega)\equiv\mathcal{L}(\omega)$ is called the
likelihood, $P(\omega)$ is some set of priors, and $P(\d)$, the so-called evidence, is effectively a
normalization constant for purposes of evaluating $\omega$. The likelihood is easily defined, and given by
Eq.~\eqref{eq:simplemodel} under the assumption that
$\n^{\mathrm{w}}_{j}$ is Gaussian distributed,
\begin{equation}
-2\ln\mathcal{L}(\omega) = \left(\d-\s^{\mathrm{tot}}(\omega)\right)^t\N_{\mathrm{wn}}^{-1}\left(\d-\s^{\mathrm{tot}}(\omega)\right).
\end{equation}
The prior is less well-defined, and we adopt in practice a combination
of informative and algorithmic priors in the \BP\ analysis (see
\citet{bp01} for an overview).

To explore this distribution by Markov Chain Monte Carlo, we use the following Gibbs sampling
chain \citep{bp01},
\begin{alignat}{11}
\g &\,\leftarrow P(\g&\,\mid &\,\d,&\, & &\,\xi_n, &\,\a^{\mathrm{1Hz}}, &\,\Dbp, &\,\a, &\,\beta, &\,C_{\ell})\label{eq:gibbs_g}\\
\n_{\mathrm{corr}} &\,\leftarrow P(\n_{\mathrm{corr}}&\,\mid &\,\d, &\,\g, &\,&\,\xi_n,
&\,\a^{\mathrm{1Hz}}, &\,\Dbp, &\,\a, &\,\beta, &\,C_{\ell})\\
\xi_n &\,\leftarrow P(\xi_n&\,\mid &\,\d, &\,\g, &\,\n_{\mathrm{corr}}, &\,
&\,\a^{\mathrm{1Hz}}, &\,\Dbp, &\,\a, &\,\beta, &\,C_{\ell})\\
\a^{\mathrm{1Hz}} &\,\leftarrow P(\a^{\mathrm{1Hz}}&\,\mid &\,\d,
&\,\g, &\,\n_{\mathrm{corr}}, &\,\xi_n, &\,
&\,\Dbp, &\,\a, &\,\beta, &\,C_{\ell})\\
\Dbp &\,\leftarrow P(\Dbp&\,\mid &\,\d, &\,\g, &\,\n_{\mathrm{corr}}, &\,\xi_n,
&\,\a^{\mathrm{1Hz}}, &\,&\,\a, &\,\beta, &\,C_{\ell})\\
\beta &\,\leftarrow P(\beta&\,\mid &\,\d, &\,\g, &\,\n_{\mathrm{corr}}, &\,\xi_n,
&\,\a^{\mathrm{1Hz}}, &\,\Dbp, & &\,&\,C_{\ell})\\
\a &\,\leftarrow P(\a&\,\mid &\,\d, &\,\g, &\,\n_{\mathrm{corr}}, &\,\xi_n,
&\,\a^{\mathrm{1Hz}}, &\,\Dbp, &\,&\,\beta, &\,C_{\ell})\\
C_{\ell} &\,\leftarrow P(C_{\ell}&\,\mid &\,\d, &\,\g, &\,\n_{\mathrm{corr}}, &\,\xi_n,
&\,\a^{\mathrm{1Hz}}, &\,\Dbp, &\,\a, &\,\beta&\,\phantom{C_{\ell}})&,\label{eq:gibbs_cl}
\end{alignat}
where the symbol $\leftarrow$ denotes setting the variable on the
left-hand side equal to a sample from the distribution on the
right-hand side.

\section{Bayesian versus frequentist simulations}
\label{sec:frequentist}

\begin{figure*}[t]
  \begin{center}
    \includegraphics[width=\linewidth]{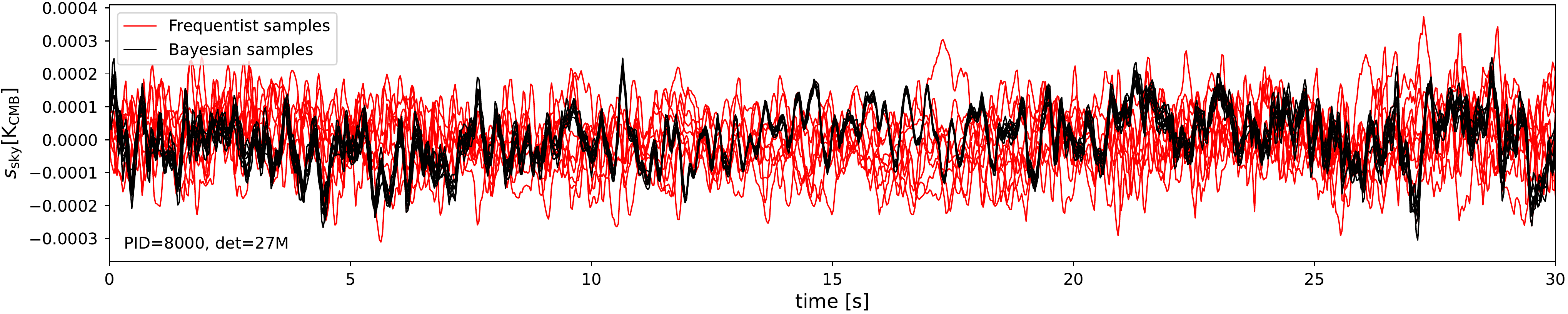}\\
    \includegraphics[width=\linewidth]{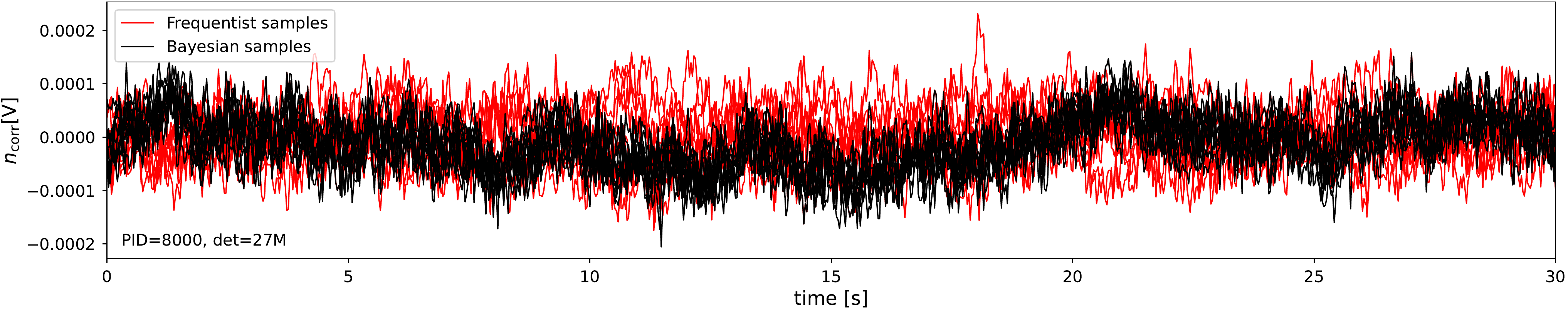}
    \caption{Comparison of ten frequentist (red) and ten Bayesian
      (black) simulations in time-domain. Each line represents one
      independent realization of the respective type. The top panel
      shows sky model (i.e., CMB) simulations and the bottom panel
      shows correlated noise simulations.}
    \label{fig:freq_vs_bayes}
  \end{center}
\end{figure*}

End-to-end TOD simulations have become the \textit{de-facto} industry standard
for producing robust error estimates for high-precision experiments
\citep[e.g.,][]{planck2014-a14}, and the data model defined in
Eqs.~\eqref{eq:todmodel}--\eqref{eq:1fmodel_lognorm} represents a
succinct simulation recipe for producing such simulations: If $\omega$
is assumed to be perfectly known, then these equations can be
evaluated in a forward manner without the need for parameter
estimation or inversion algorithms, and the only stochastic terms are
the correlated and white noise, both of which can be easily generated
by a combination of standard random Gaussian number generators and
Fourier transforms. 

However, in practice $\omega$ is of course not perfectly known, and
precisely how $\omega$ is specified has direct and strong implications
regarding what the resulting simulations can inform the user about;
for an example of this within the context of \Planck\ LFI, see
\citet{bp10}. In short, the key discriminator is whether $\omega$ is
defined using real observed data (and in practice drawn from the
posterior distribution, $P(\omega\mid\d)$) or whether it is drawn from a
data-independent hyper-distribution, for instance informed by
theoretical models and/or ground-based laboratory measurements. 
We will refer to these two approaches as ``Bayesian'' and ``frequentist'' 
respectively, indicating whether or not they condition on the true
data in question.

We note that both Bayesian and frequentist simulations specifically
refer to time-ordered data in the current paper, not pixelized maps
or higher-level products. That is, we distinguish between simulation
pipelines, which transform $\omega$ into timelines, and analysis
pipelines, which transform timelines into higher-ordered products,
such as maps and power spectra. 

\subsection{Bayesian versus frequentist statistics}

Before comparing the two simulation types through a few worked
examples, it is useful to recall the fundamental difference between
Bayesian and frequentist statistics, which may be summarized as
follows: In frequentist statistics, the model $\mathcal{M}$ and its
parameters $\omega$ are considered to be fixed and known, while the
data $\d$ are considered to be the main uncertain quantity. In
Bayesian statistics, on the other hand, $\d$ is assumed to be
perfectly known, and essentially defined by a list of numbers recorded
by a measuring device, while $\omega$ is assumed to be the main
unknown quantity.

\begin{figure*}[p]
  \center
  \includegraphics[width=0.33\linewidth]{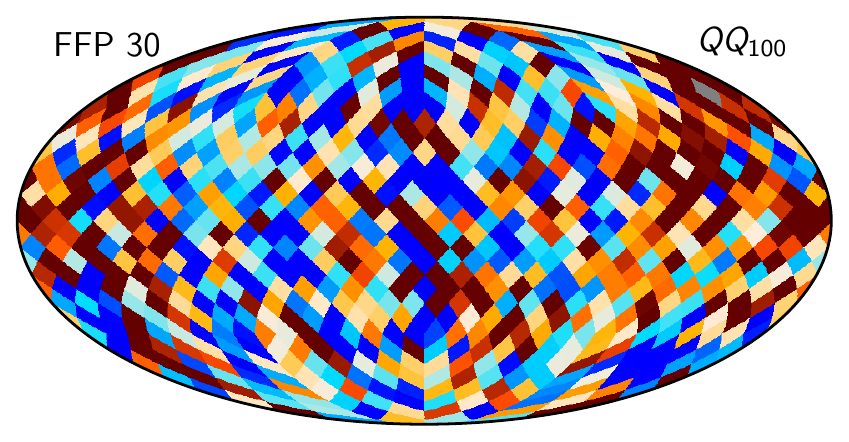}
  \includegraphics[width=0.33\linewidth]{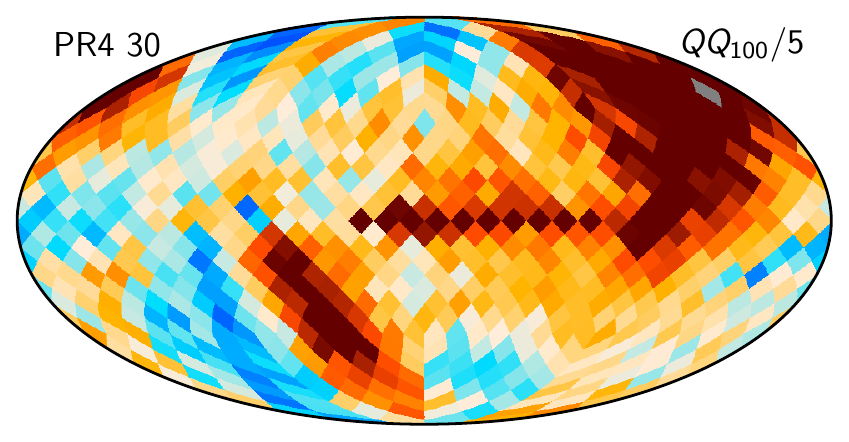}
  \includegraphics[width=0.33\linewidth]{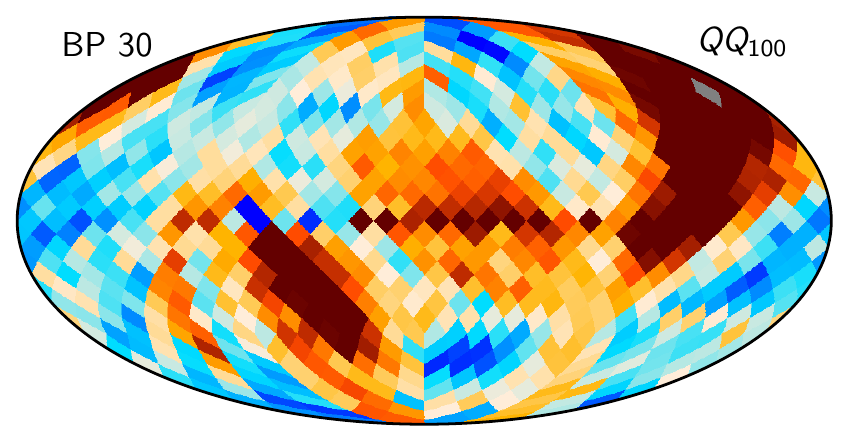}\\
  \includegraphics[width=0.33\linewidth]{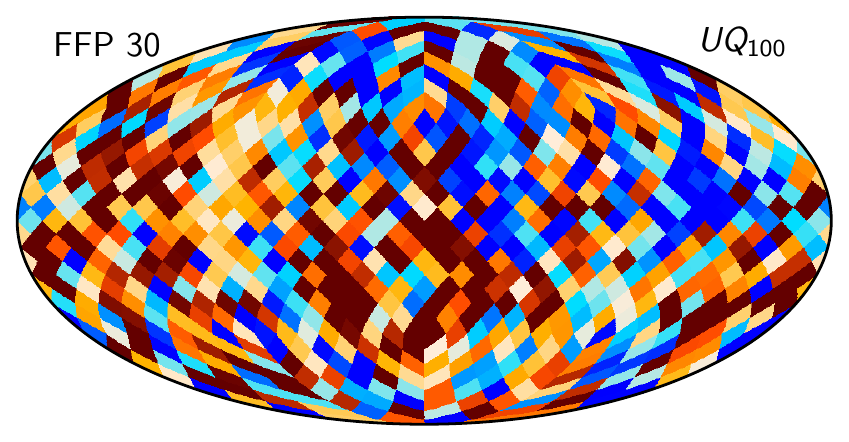}
  \includegraphics[width=0.33\linewidth]{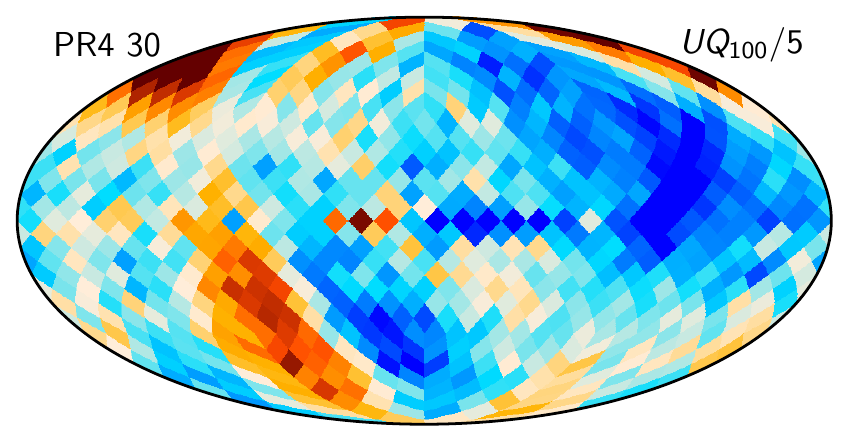}
  \includegraphics[width=0.33\linewidth]{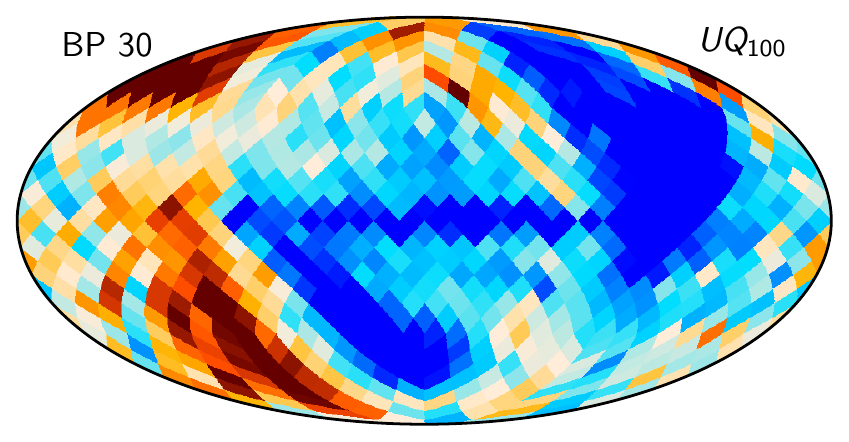}\\\vspace*{5mm}
  \includegraphics[width=0.33\linewidth]{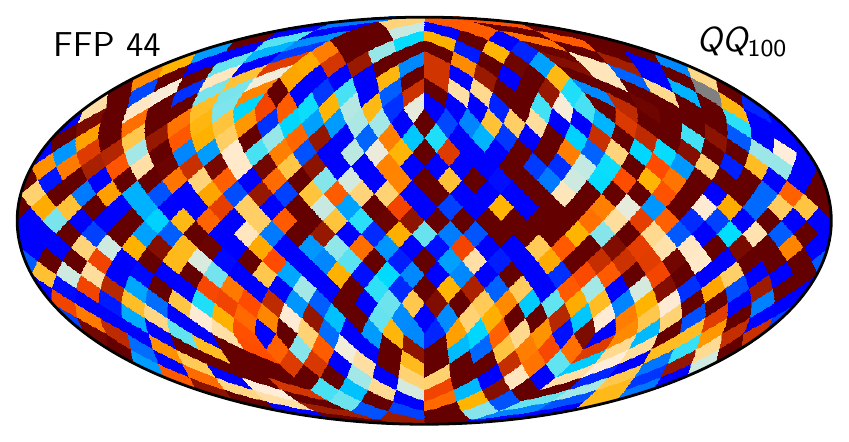}
  \includegraphics[width=0.33\linewidth]{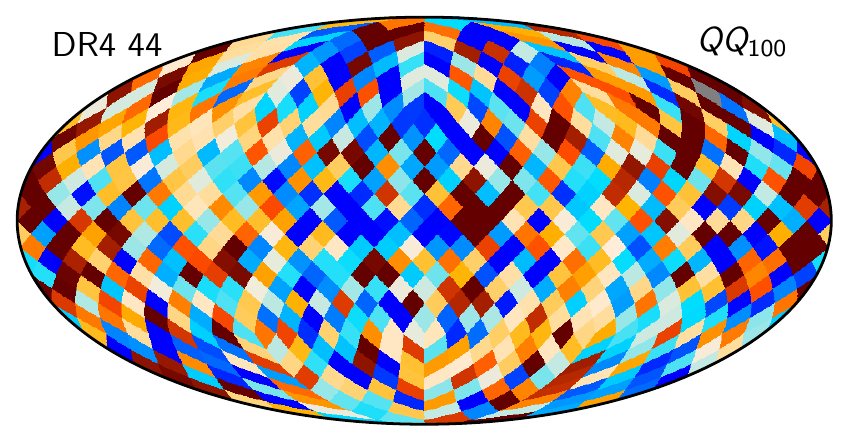}
  \includegraphics[width=0.33\linewidth]{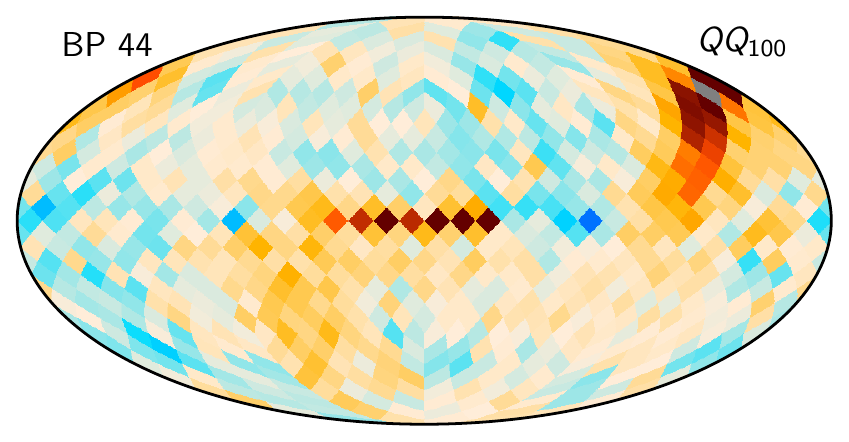}\\
  \includegraphics[width=0.33\linewidth]{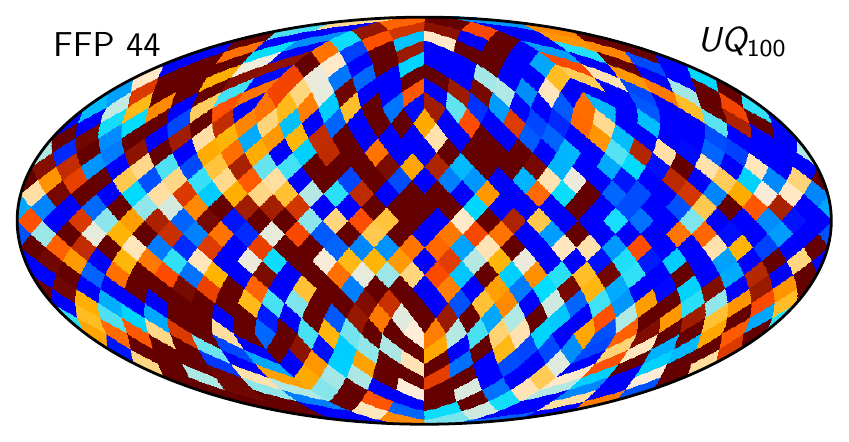}
  \includegraphics[width=0.33\linewidth]{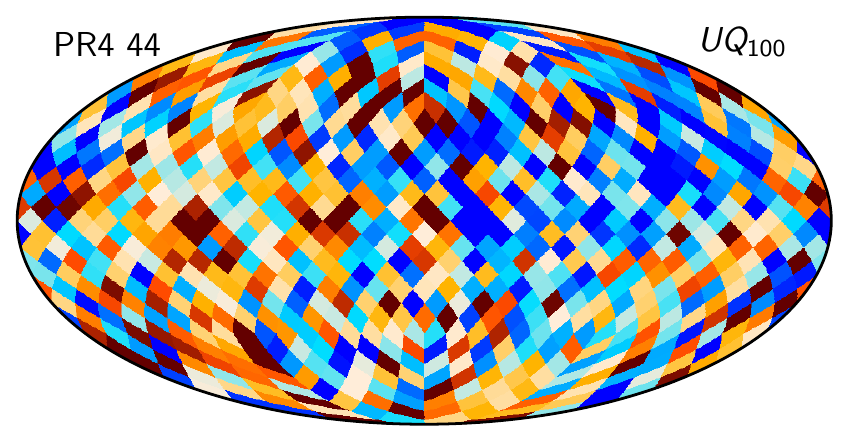}
  \includegraphics[width=0.33\linewidth]{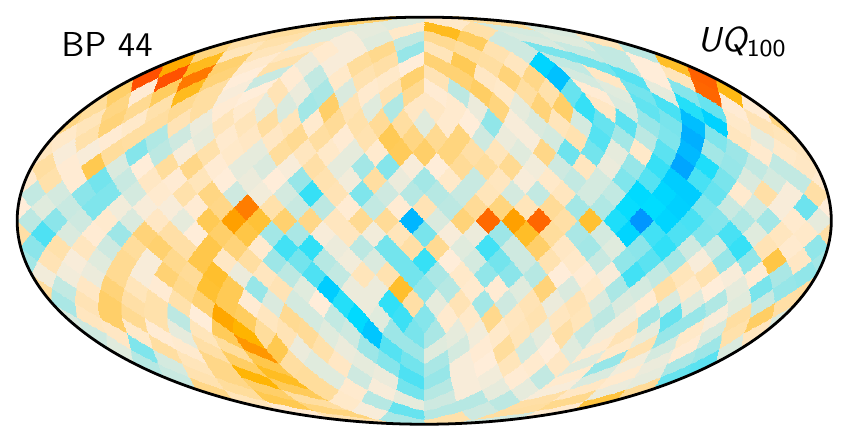}\\\vspace*{5mm}
  \includegraphics[width=0.33\linewidth]{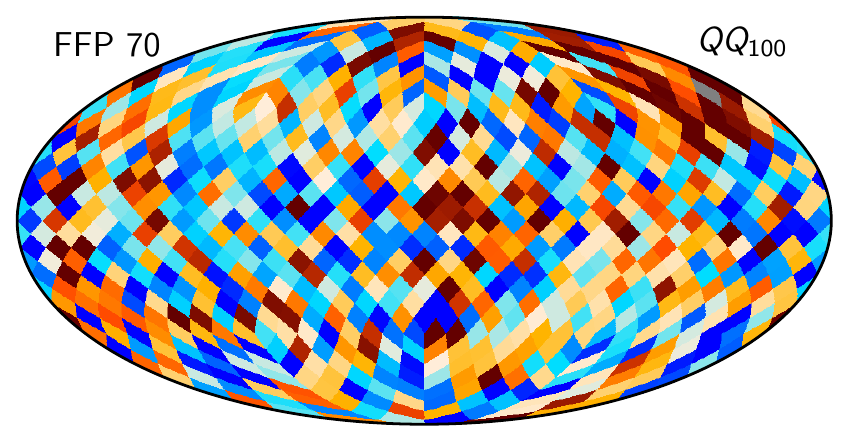}
  \includegraphics[width=0.33\linewidth]{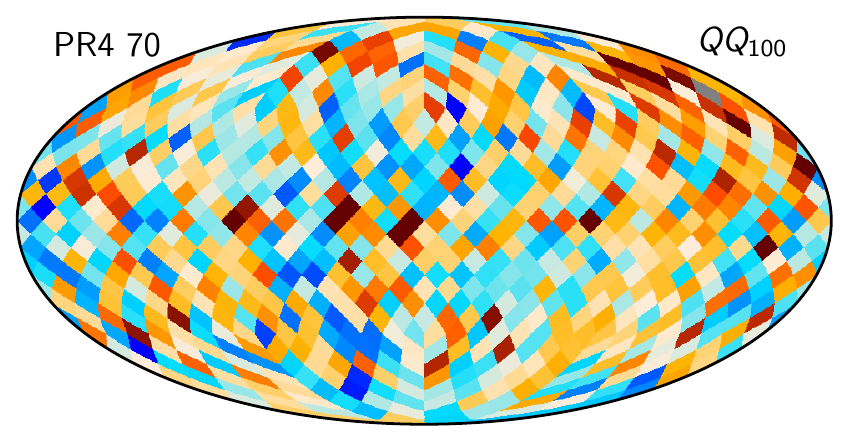}
  \includegraphics[width=0.33\linewidth]{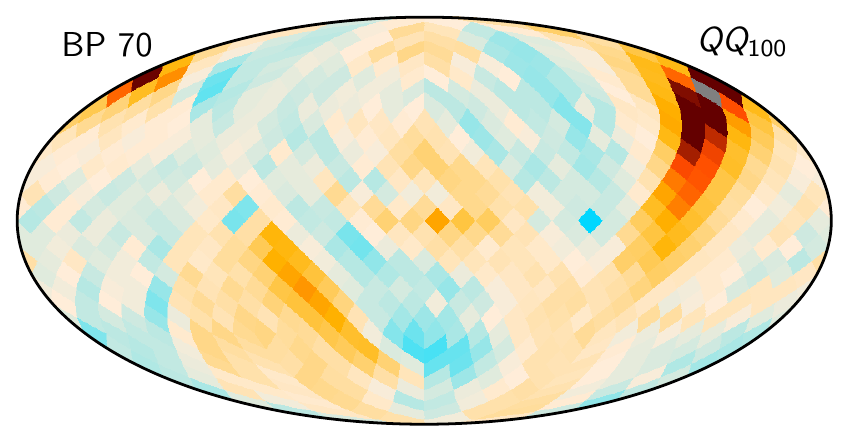}\\
  \includegraphics[width=0.33\linewidth]{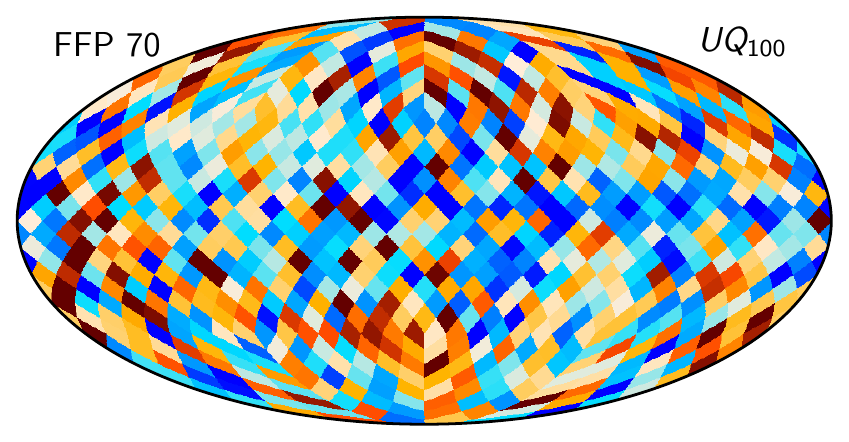}  
  \includegraphics[width=0.33\linewidth]{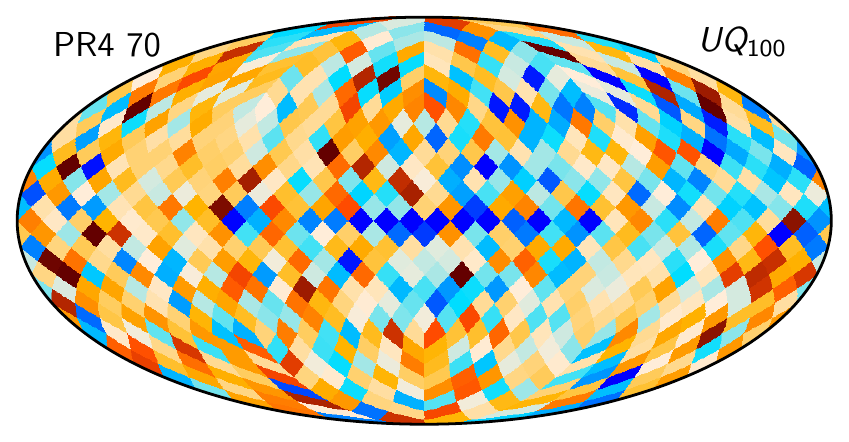}
  \includegraphics[width=0.33\linewidth]{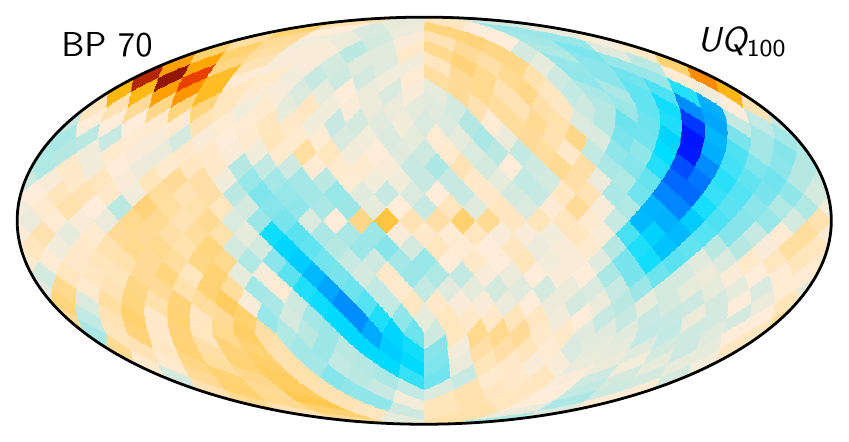}\\
  \includegraphics[width=0.35\linewidth]{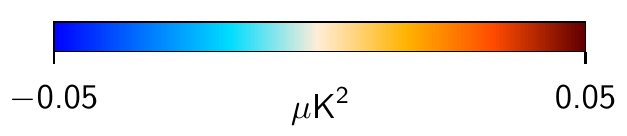}
  \caption{Single column of the low-resolution 30 (\emph{top
      section}), 44 (\emph{middle section}), and 70\,GHz (\emph{bottom
      section}) frequency channel covariance matrix, as estimated from
    300 LFI DPC FFP10 frequentist simulations (\emph{left column}); from
    300 PR4 frequentist simulations (\emph{middle column}); and from
    3200 \BP\ Bayesian simulations (\emph{right column}). The selected
    column corresponds to the Stokes $Q$ pixel number 100 marked in gray, which
    is located in the top right quadrant. All 
    covariance matrices are constructed at $N_{\mathrm{side}}=8$. 
    Note that the \Planck\ PR4 30\,GHz covariance slice
    has been divided by a factor of 5, and it is therefore even stronger
    than the color scale naively implies.}
  \label{fig:ncov}
\end{figure*}

This difference has important consequences for how each framework
typically approaches statistical inference, and which questions they
are most suited to answer. This is perhaps most easily illustrated
through their most typical mode of operations. First, the classical
frequentist approach to statistical inference is to construct an
ensemble of simulated data sets, $\d_i$, each with parameters drawn
independently from $\mathcal{M}(\omega)$. The next step is to define
some statistic, $\gamma(\d_i): \mathbb{R}^N \rightarrow \mathbb{R}$,
that isolates and highlights the important piece of information that
the user is interested in; widely used CMB examples include $\chi^2$
statistics, angular power spectrum statistics, or non-Gaussianity
statistics. Finally, one computes $\gamma$ both for the simulations
and the actual data, and determines the relative frequency for which
$\gamma(\d_{\mathrm{real}}) < \gamma(\d_i)$, which is often called the
$p$-value or ``probability-to-exceed'' (PTE). Values between, say,
0.025 and 0.975 are taken to suggest that the data are consistent with
the model, while more extreme values indicate a discrepancy.

Given this prescription, it is clear that the frequentist approach is
particularly suited for model testing applications; it intrinsically
and directly addresses the question of whether the data are consistent
with the model. As such it has been widely used in the CMB field for
instance for studies of non-Gaussianity and isotropy. In this case,
the null-hypothesis is easy to specify, namely that the universe is
isotropic and homogeneous, and filled with Gaussian random
fluctuations drawn from a $\Lambda$CDM universe with given
parameters. Establishing some statistic that shows that the observed
data are inconsistent with this hypothesis would constitute evidence
of new physics, and is as such a high-priority scientific target.

In contrast, Bayesian statistics takes a fundamentally different
approach to statistical inference. In this case, we consider $\omega$
to be a stochastic and unknown quantity, and want to understand how
the observed data constrains $\omega$. The most succinct summary of
this is the posterior probability distribution itself, $P(\omega\mid\d)$,
and the starting point for this framework is therefore Bayes' theorem
as given in Eq.~\eqref{eq:jointpost}. The majority of applications of
modern Bayesian statistics thus simply amounts to mapping out
$P(\omega\mid\d)$ as a function of $\omega$ by any means necessary.

At the same time, it is important to note that the likelihood
$\mathcal{L}(\omega) = P(\d\mid\omega)$ on the right-hand side of
Eq.~\eqref{eq:jointpost} is a fully classical frequentist statistic,
in which $\omega$ is assumed to be perfectly known, and the data are
uncertain. Still, it is important to note that the free parameter in
$\mathcal{L}(\omega)$ is indeed $\omega$, not $\d$, and $\mathcal{L}$
itself is really just a frequentist statistic that measures the
overall goodness-of-fit between the data and the model. This statistic
may then be used to estimate $\omega$ within a strictly frequentist
framework; one popular example of this within the CMB field are
so-called profile likelihoods. 

Likewise, the Bayesian approach is also able to address the model
selection problem, and this is most typically done using the evidence
factor, $P(\d)$, in Eq.~\ref{eq:jointpost}. The importance of this
factor becomes obvious when explicitly acknowledging that all involved
probability distributions in Eq.~\eqref{eq:jointpost} actually depend
on the overall model $\mathcal{M}$, and not only the individual
parameter values,
\begin{equation}
  P(\omega\mid \d, \mathcal{M}) = \frac{P(\d\mid \omega, \mathcal{M})P(\omega\mid\mathcal{M})}{P(\d\mid\mathcal{M})} 
  \label{eq:bayes}.
\end{equation}
Mathematically, $P(\d\mid\mathcal{M})$ is simply given by the average
likelihood integrated over all allowed parameter values, and classical
Bayesian model selection between models $\mathcal{M}_1$ and
$\mathcal{M}_2$ proceeds simply by evaluating
$P(\d\mid\mathcal{M}_1)/P(\d\mid\mathcal{M}_2)$; the model with the higher
evidence is preferred.

In summary, the foundational assumptions underlying frequentist and
Bayesian methods are different and complementary, and they
fundamentally address different questions. Frequentist statistics are
ideally suited to address model testing problems (e.g., ``is the
observed CMB sky Gaussian and isotropic?''), while Bayesian statistics
are ideally suited to address parameter estimation problems (e.g.,
``what are the best-fit $\Lambda$CDM parameters?''). At the same time,
this dichotomy is by no means absolute, and either framework is fully
capable of addressing both types of questions if they are carefully posed.

\subsection{Constrained versus random input parameters in CMB simulations}

We now return to the issue raised in the introduction to this section,
namely how to properly choose $\omega$ for CMB inference based on
end-to-end simulations. As discussed by \citet{bp10}, essentially all
CMB analysis pipelines prior to \BP\ have adopted a mixture of
data-constrained and data-independent parameters for this purpose. Key
examples of the former are the CMB Solar dipole and Galactic
foregrounds, both of which are strongly informed by real
measurements. Correspondingly, classical examples of the latter are
CMB fluctuations, which are typically drawn as Gaussian realizations
from a $\Lambda$CDM power spectrum, and instrumental noise, which is
often based on laboratory measurements. In our notation, these
simulations qualify thus neither as pure Bayesian nor pure
frequentist, but rather as a mixture of the two.

In contrast, each sample of $\omega$ produced by the \BP\ Gibbs chain
summarized in Eqs.~\eqref{eq:gibbs_g}--\eqref{eq:gibbs_cl} represents
one possible simulated realization in which \emph{all} sub-parameters
in $\omega$ are determined exclusively by the real posterior
distribution; not only the CMB dipole and Galactic model, but also
those parameters that are traditionally chosen from external sources
in classical pipelines, such as the CMB anisotropies and the
specific noise realization.

The difference between these two types of simulation inputs is
illustrated in Fig.~\ref{fig:freq_vs_bayes} which compares ten
independent frequentist time-domain realizations (red curves) with ten
independent Bayesian realizations (black curves). The top and bottom
panels show the correlated noise $n_\mathrm{corr}$ and the sky model
$s_\mathrm{sky}$, respectively, both plotted as a function of
time. Starting with the frequentist simulations, we see that these are
entirely uncorrelated between realizations, and scatter randomly with
some model-specific mean and variance. In particular, the frequentist
simulations include so-called cosmic variance, i.e., independent
realizations have different CMB and noise amplitudes and phases, even
if they are drawn from the same underlying stochastic model. In
contrast, Bayesian simulations do not include cosmic variance, but
rather focus exclusively on structures in the real data. For the sky
signal component shown in the top panel of
Fig.~\ref{fig:freq_vs_bayes}, this is seen in terms of two different
aspects. First, the structure of all ten realizations follow very
closely the same overall structure, and this is defined by the
specific CMB pattern of the real sky. However, they also explicitly
account for the uncertainty in the sky value at each pixel, and this
is seen by the varying width of the black band; in the middle of the
plot, the width is small, and this implies that the sky has been well
measured here (due to deep scanning), while along the edges of the
plot the width is larger, and this implies that the sky as been less
well measured. The variation between Bayesian simulations thus
directly quantify the uncertainty of the true data. Intuitively
speaking, this point may be summarized as follows: Uncertainties
measured by frequentist simulations quantify the expected variations
as observed with a \emph{random} instrument in a \emph{random}
universe, while Bayesian simulations quantify the expected variations
of the \emph{real} instrument in the \emph{real} universe.

These intuitive differences translate directly into both qualitatively
and quantitatively different ensemble properties for the resulting
simulations, and correspondingly also into different resulting error
estimates. As a real-world illustration of this, Fig.~\ref{fig:ncov}
shows slices through the empirical low-resolution polarization
covariance matrix computed for each of the three \Planck\ LFI
frequency channels using three different generations of LFI
simulations, namely (from left to right columns) \Planck\ 2018
\citep{planck2016-l02}, \Planck\ PR4 \citep{planck2020-LVII}, and
\BP\ \citep{bp01}. Row sections show results for the 30, 44, and
70\,GHz channels, respectively, and within each section the two rows
show the $QQ$ and $UQ$ segments of the full matrix, sliced through
Stokes $Q$ pixel number 100, marked in gray in the upper right
quadrant. Each covariance matrix is computed by first downgrading each
simulation to a HEALPix\footnote{\url{https://healpix.jpl.nasa.gov}}
\citep{gorski2005} resolution of $N_{\mathrm{side}}=8$, and averaging
the outer product over all available realizations; see
\citet{bp10,bp11} for further details. Effectively, these matrices
visually summarize the map-space uncertainty estimates predicted by
each simulation set.

Starting with the \Planck\ 2018 simulations, the most striking
observation is that these empirical matrices are very noisy for all
three frequency channels. This is partly a reflection of the fact that
only 300 simulations were actually constructed, and this leads to a
high Monte Carlo uncertainty. However, it is also a reflection of the
fact that these simulations are largely frequentist based, with both random
CMB and noise properties. Furthermore, the gains that were
assumed when generating these simulations exhibited significantly less
structure than the real observations. In sum, there are relatively
little common structures between the various realizations, either from
the astrophysical sky, the instrumental noise, or the gain, and the 
corresponding covariance structures are therefore weak. Visually
speaking, perhaps the most notable feature is a positive correlation
from correlated noise along the scanning direction that passes through
the sliced pixel seen in the upper right quadrant, but these are
significantly obscured by Monte Carlo uncertainties.

Proceeding to the \Planck\ PR4 simulations summarized in the middle
column, we now see very strong coherent structures for the 30\,GHz
channel, while the 44 and 70\,GHz
channels behave similarly to the 2018 case. The
explanation for this qualitative difference is the \Planck\ PR4
calibration algorithm; in this pipeline, the 30\,GHz channel is
calibrated independently without the use of supporting priors, while
the 44 and 70\,GHz channels are calibrated by using the 30\,GHz
channel as a polarized foreground prior. The net effect of this
independent calibration procedure is a very high calibration
uncertainty for the 30\,GHz channel, and these couple directly to the
true CMB dipole, which is kept fixed between all simulations. The
result is the familiar large-scale pattern seen in this figure, which
has been highlighted by several previous analyses as a particularly
difficult mode to observe with
\Planck\ \citep[e.g.,][]{planck2016-l02,bp07,bp17}.

Turning to the \BP\ simulations summarized in the right column, we now
see coherent and signal-dominated structures across the full sky in
all frequency channels. A part of this is simply due to more
realizations than for the other two pipelines --- in this case 3200
--- but even more importantly is the fact that the simulations are now
entirely data-driven. That is, they correspond to the black curves in
Fig.~\ref{fig:freq_vs_bayes}, while the previous pipelines correspond
to the red curves. In practice, this has two main effects. First, it
implies that the total parameter volume that needs to be explored by
Monte Carlo sampling is intrinsically smaller, simply because the
posterior distribution does not include cosmic variance; the
simulations only need to describe \emph{our} instrument and universe,
not \emph{any} instrument and universe, and this is a much smaller
sub-set. Second, and even more importantly, the Bayesian simulations
account naturally for \emph{non-linearity} between the various
parameters, and these are very often the dominant contributions in
these distributions. As a concrete example, if the gain happens to
scatter either high or low during a given time period, then the total
uncertainty estimate will be particularly sensitive to the CMB dipole
during the same time period, and it will excite a correlation
structure in these plots that is intimately connected to the satellite
scanning strategy. Thus, if one chooses a gain profile that is
independent of other parameters, then those real uncertainties will
not be properly accounted for in the simulation set: Intuitively
speaking, the hot and cold spots in the covariance matrices shown in
Fig.~\ref{fig:ncov} will either appear in the wrong places, or be
suppressed when averaging over independent realizations. In general,
specifying the instrumental model at a sufficiently realistic level
represents a real challenge for frequentist simulations, and great
care is required in order to capture the full error budget. This task
is considerably simplified in the Bayesian approach, as each
instrumental parameter is defined directly from the data themselves.

\section{Simulation specification}
\label{sec:simulation}

\begin{figure}[t]
  \center
  \includegraphics[width = \linewidth]{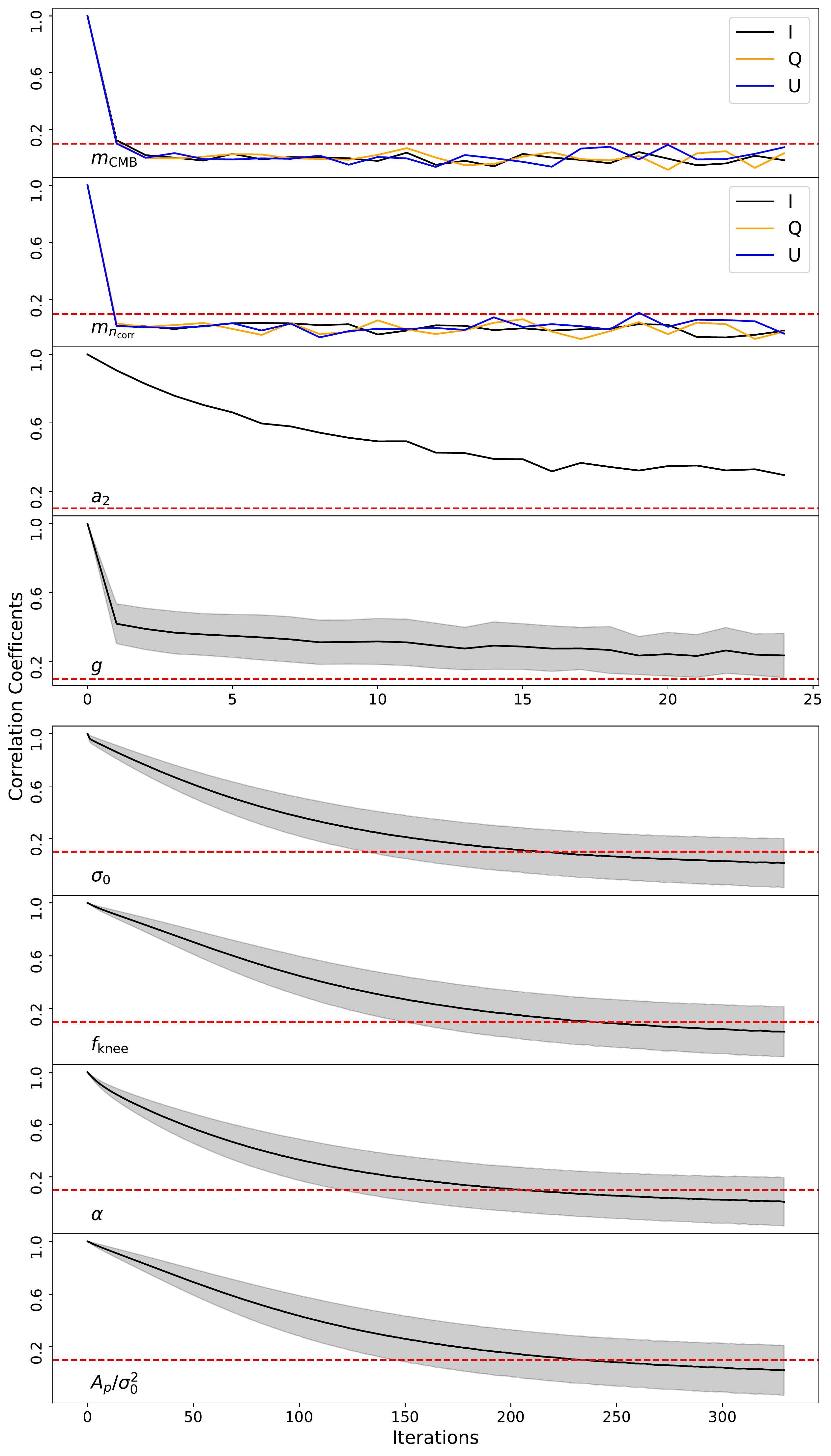}
  \caption{Auto-correlation function, $\rho$, for selected parameters
    in the model, as estimated from a single chain with 10\,000
    samples. From top to bottom, the various panels show 1) one pixel
    value of the CMB component map $m_\mathrm{CMB}$; 2) one pixel of
    the correlated noise map $m_{n_\mathrm{corr}}$; 3) the temperature
    quadrupole moment, $a_{2,0}$; 4) the PID-averaged total gain $g$;
    and 5--8) the PID-averaged noise PSD parameters $\sigma_0$,
    $f_{\mathrm{knee}}$, $\alpha$, and $A_{\mathrm{p}}/\sigma_0$. In
    panels with multiple lines, the various colors show Stokes $T$,
    $Q$, and $U$ parameters. In panels with gray bands, the black line
    shows results averaged over all PIDs, and the band shows the
    $1\,\sigma$ variation among PIDs. The dashed red line marks a
    correlation coefficient of 0.1, which is used to define the
    typical correlation length of each parameter.
  }
  \label{fig:spectral_burnin_corrlen}
\end{figure}

Returning to the data model summary in Sect.~\ref{sec:pipeline}, we
note that the \commanderthree\ code described by \citet{bp03}, and
used by the \BP\ project to perform Bayesian end-to-end analysis of
the \Planck\ LFI data, is able to produce both frequentist and
Bayesian simulations essentially without modifications; the only
question is whether the parameters used to generate the TOD, $\omega$,
are drawn from the posterior distribution, or whether they are
selected from a data-independent hyper-distribution. Choosing which
type of simulations to generate is thus only a matter of selecting
proper initialization values in the \commanderthree\ parameter file.

In this paper, we demonstrate the frequentist mode of operation by generating a set of classical frequentist simulations with
\commanderthree, and we then use these to validate the novel
low-level processing algorithms introduced by \citet{bp02,bp06,bp07}
for mapmaking, correlated noise estimation, and gain estimation,
respectively. 

We note that the original \BP\ analysis required 670\,000 CPU-hours to
generate 4000 full Gibbs samples for the full LFI dataset, which took
about three months of runtime to complete. In the current paper, we
are primarily interested in validating the low-level algorithms
themselves, and we therefore choose to consider only one year of
30\,GHz observations in the following (corresponding to about 10,000
\Planck\ pointing periods (PIDs), each lasting for about one hour;
\citealp{planck2013-p01}), rather than the full LFI dataset, and this
reduces the computational cost from 169 to 2.5\,CPU-hours per Gibbs
sample \citep{bp03}. As a result, we are able to produce individual
chains with 10\,000 samples within a matter of days, rather than
months or years, which is useful for convergence analyses. This also
reduces the total volume of the TOD themselves (not including
pointing, flags, etc.) from 638\,GB to 22\,GB, and the simulations may
therefore be run on a much broader range of hardware. In fact, subsets
of the following simulations have been produced on more than ten
different computing systems all over the world, using both AMD and
Intel processors (e.g., Intel~E5-2697v2 2.7\,GHz, Intel Xeon E5-2698
2.3\,GHz, Intel~Xeon~W-2255 3.7\,GHz, AMD Ryzen~9 3950X 2.2\,GHz),
with between 128\,GB and 1.5\,TB RAM per node, and using both Intel
and GNU compilers.\footnote{The research presented in this paper was
  undertaken as a part of the Master- and PhD-level course called
  ``AST9240 -- Cosmological component separation'' in 2021 at the
  University of Oslo, and individual students produced and analyzed
  simulations in their home institutions.}

Given that we will only consider low-level processing of the 30\,GHz
channel, we simplify the data model in Eq.~\eqref{eq:todmodel} to
\begin{align}
  d_{j,t}^{\mathrm{sim}} &= g_{j,t}\P_{tp,j}
      \B^{\mathrm{symm}}_{pp',j} a^\mathrm{cmb}_{p'}  +
      \B^{\mathrm{asymm}}_{pp',j} s^{\mathrm{orb}}_{j,t} + 
      + n^{\mathrm{corr}}_{j,t} + n^{\mathrm{w}}_{j,t}\\
      &= s^{\mathrm{tot}}_{j,t} + n^{\mathrm{corr}}_{j,t} + n^{\mathrm{w}}_{j,t}.
    \label{eq:todmodel2}
\end{align}
That is, we only include one single sky component, namely the CMB, and
we ignore sub-dominant effects such as far sidelobe corrections, 1\,Hz
electronic spikes, etc. As such, this configuration provides a
test of the gain, noise estimation, and mapmaking parts of the full
algorithm, but not the component separation or cosmological
parameter estimation. 

The CMB sky realizations used in the following analysis are drawn from
the best-fit \Planck\ 2018 \LCDM\ model \citep{planck2016-l05} using the
HEALPix\footnote{\url{http://healpix.jpl.nasa.gov}} \citep{gorski2005}
\texttt{synfast} utility. All instrumental parameters are drawn from
different realizations of the \BP\ ensemble presented in \citet{bp01},
and these are taken as true input values in the following. 

For the noise terms, we draw a random Gaussian realization of
$n_{j,t}= n^{\mathrm{corr}}_{j,t} + n^{\mathrm{w}}_{j,t}$ with the
noise PSD model given in Eq.~\eqref{eq:1fmodel_lognorm}. This is done
independently for each \Planck\ pointing ID (PID), and the noise PSD
parameters are thus varying in time with the same structure as the real
observations.

\section{Validation of low-level processing algorithms}
\label{sec:validation}

To validate the noise and gain estimation and mapmaking steps in
\commanderthree\,, we analyze the frequentist simulations described
above with the same Bayesian framework as used for the main
\BP\ processing, and compare the output marginal posterior
distributions with the known true inputs. To quantify both biases and
the accuracy of the uncertainty estimates, we adopt the following
normalized residual,
\begin{equation}
\delta_\omega = \frac{\mu_{\mathrm{\omega}} - \omega^{\mathrm{in}}}{\sigma_{\omega}},
\end{equation}
where $\mu_{\omega}$ and $\sigma_{\omega}$ are the posterior mean and
standard deviation for parameter $\omega$. For a truly Gaussian
posterior distribution with no bias and perfect uncertainty
estimation, this quantity should be distributed according to a
standard normal distribution with zero mean and unit variance,
$N(0,1)$, while a non-zero value of $\delta$ indicates a bias measured
in units of $\sigma$. It is of course important to note that the full
data model in Eq.~\eqref{eq:todmodel} is highly non-linear due to the
presence of the gain; therefore, the deviations from $N(0,1)$ at some level 
are fully expected, in particular for signal-dominated
quantities. Still, we find that $\delta$ serves as a useful quality
monitor.

Unless otherwise noted, the main results presented in the following
are derived from a single Markov chain comprising 10\,000
samples. Where useful for convergence and mixing assessment, we will
also use shorter and independent chains, typically with 1000 samples
in each chain.

\subsection{Markov auto-correlations}

We are also interested in studying the statistical properties of
individual Markov chains in terms of correlation lengths,
degeneracies, and convergence. We define the Markov chain
auto-correlation for a given chain as
\begin{equation}
\rho_{\omega}(\Delta) = \left<\left(\frac{\omega^i-\mu_{\omega}}{\sigma_\omega}\right)
    \left(\frac{\omega^{i+\Delta}-\mu_{\omega}}{\sigma_\omega}\right) \right>,
\end{equation}
where $i$ denotes Gibbs sample number, and $\Delta$ is a chain lag parameter which denotes sample separation.

Figure~\ref{fig:spectral_burnin_corrlen} shows the auto-correlation for
a typical set of parameters. The top four panels show (1) a single CMB
map pixel (in $T$, $Q$, and $U$); (2) a single correlated noise map
pixel (in $T$, $Q$, and $U$); (3) the CMB temperature quadrupole
moment, $a_{2,0}$; and (4) the gain for a single PID. These all have
relatively short correlation lengths, which indicates that we are
likely to produce robust results for these parameters.

\begin{figure*}[t]
  \center
  \includegraphics[width = \linewidth]{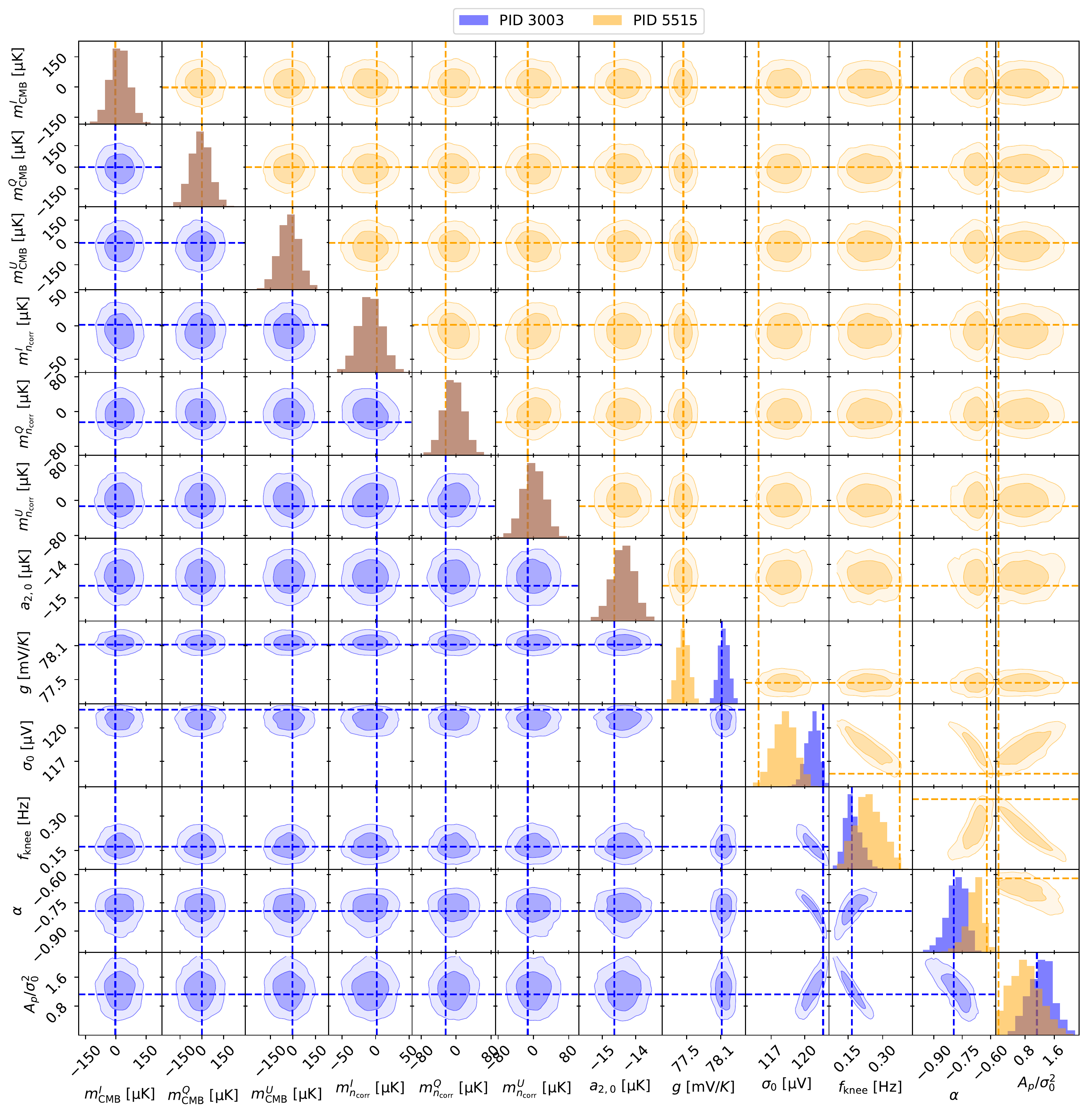}
  \caption{Recovered posterior distributions for a selected set of parameters from two PIDs and detectors. The contours indicate 68 and 95\,\% confidence regions, while the dashed lines (in the respective color of the contours) show the true input value of each of the PIDs. The contours below~(blue) and above~(orange) the diagonal correspond to PIDs 3003 and 5515, respectively. From left to right along the horizontal axis, columns show (1)--(3) one arbitrary CMB map pixel in Stokes $I$, $Q$, and $U$; (4)--(6) correlated noise for the same pixel and Stokes parameters; (7) the CMB intensity quadrupole amplitude $a_{2,0}$; (8) gain $g$; and (9)--(12) the four correlated noise parameters, $\xi^n = \{\sigma_0,f_\mathrm{knee},\alpha,A_\mathrm{p}\}$. Note that the one-dimensional histograms of the first seven parameters are completely overlapping since these parameters are independent of PID.
}
  \label{fig:noise_corner}
\end{figure*}

\begin{figure}[t]
    \includegraphics[width=\linewidth]{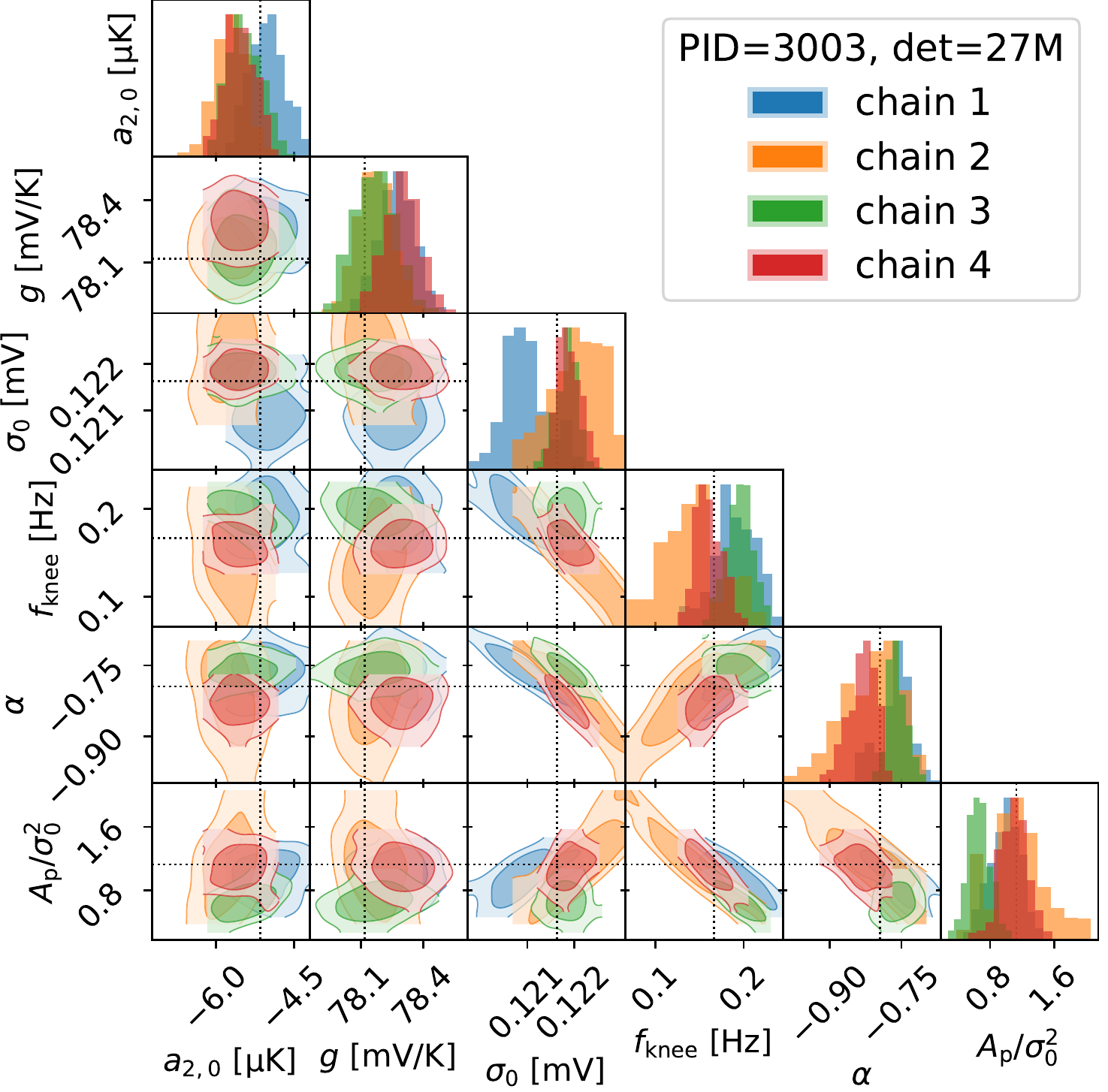}\\
    \caption{Comparison of partial posteriors distributions from
      multiple short chains for the quadrupole amplitude~$a_{2,0}$, the
      gain~$g$, white noise level~$\sigma_0$, knee
      frequency~$f_\mathrm{knee}$, correlated noise spectral
      index~$\alpha$, and log-normal noise
      amplitude~$A_\mathrm{p}$.
      Each chain consists of 1000 samples.
      The posterior contours only span the range of the underlying samples, wherefore some are not closed.}
    \label{fig:triangle_chain_comparison}

    \begin{center}
        \includegraphics[width=\linewidth]{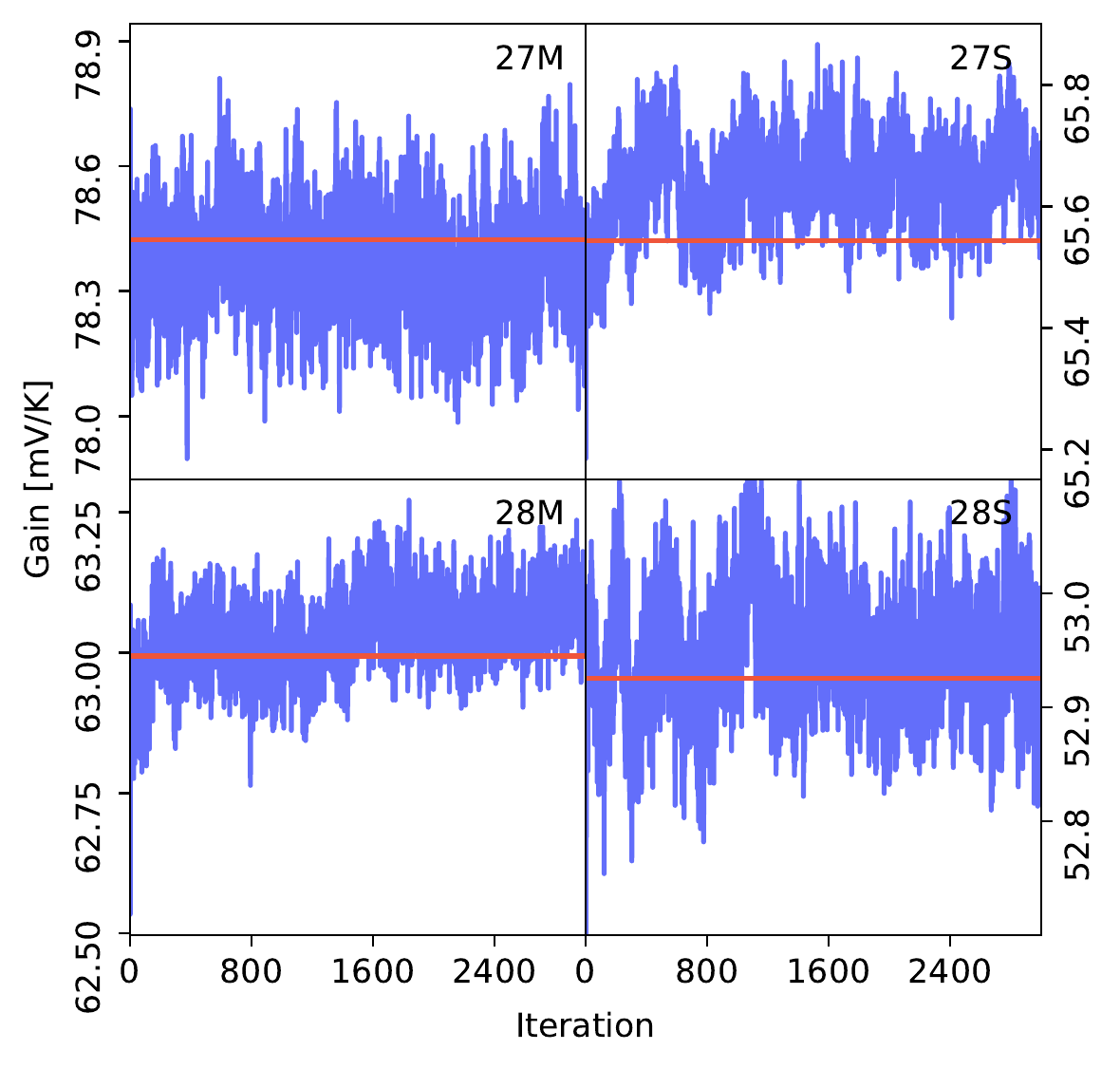}
        \caption{Trace plots of gain values as a function of chain iteration (blue) compared to their input values (red) for selected PIDs, in order from left to right: 349, 9847, 4298 and 1993.}
        \label{fig:trace_gain}
    \end{center}

\end{figure}

\begin{figure}[t]
    \begin{center}
        \includegraphics[width=\linewidth]{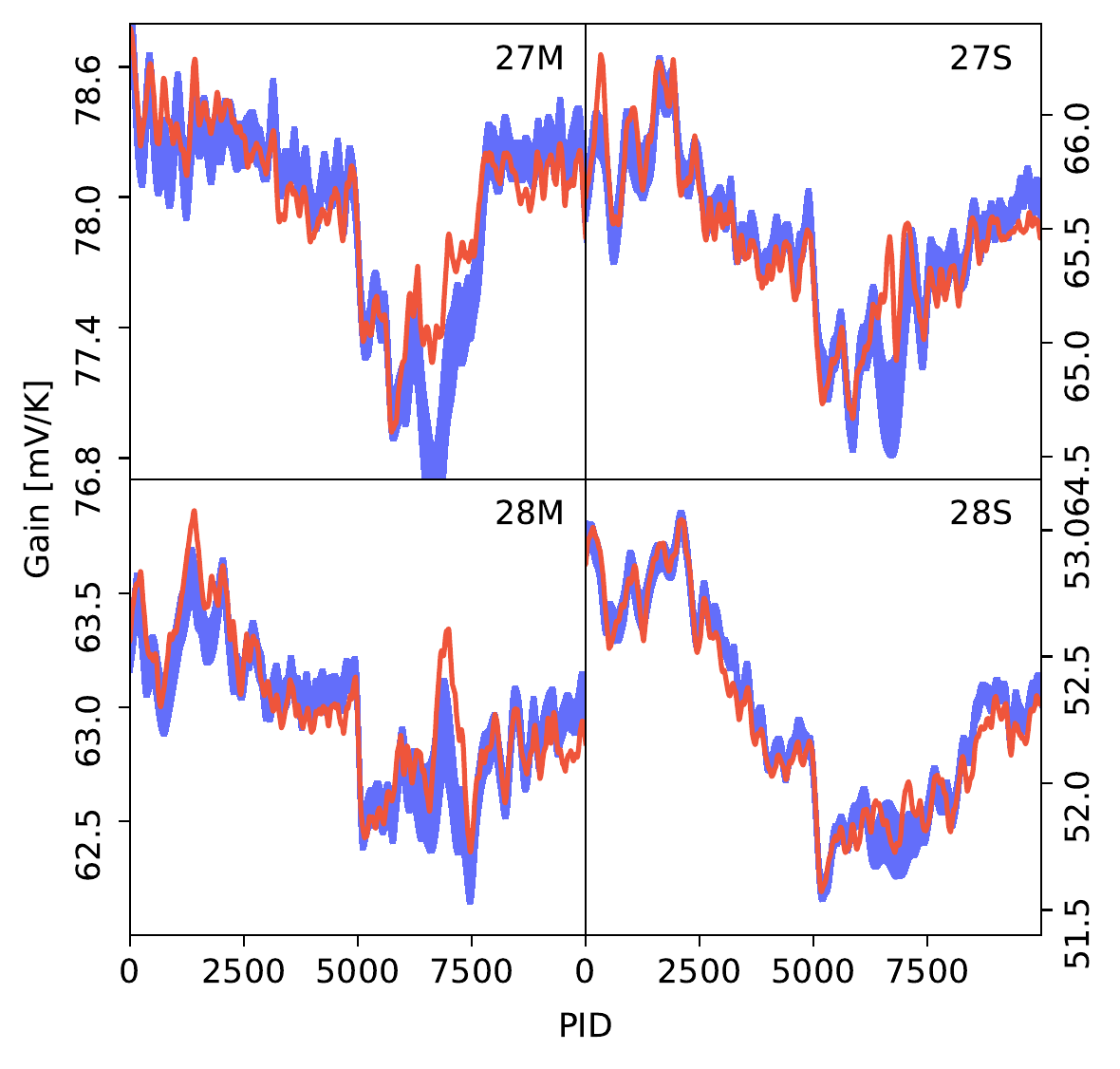}
        \caption{The input gain values (red) over-plotted on the output gain values (blue). The width of the blue line indicates the sample standard deviation of the PID in question.}            \label{fig:band_gain}

    \end{center}


    \begin{center}
        \includegraphics[width=\linewidth]{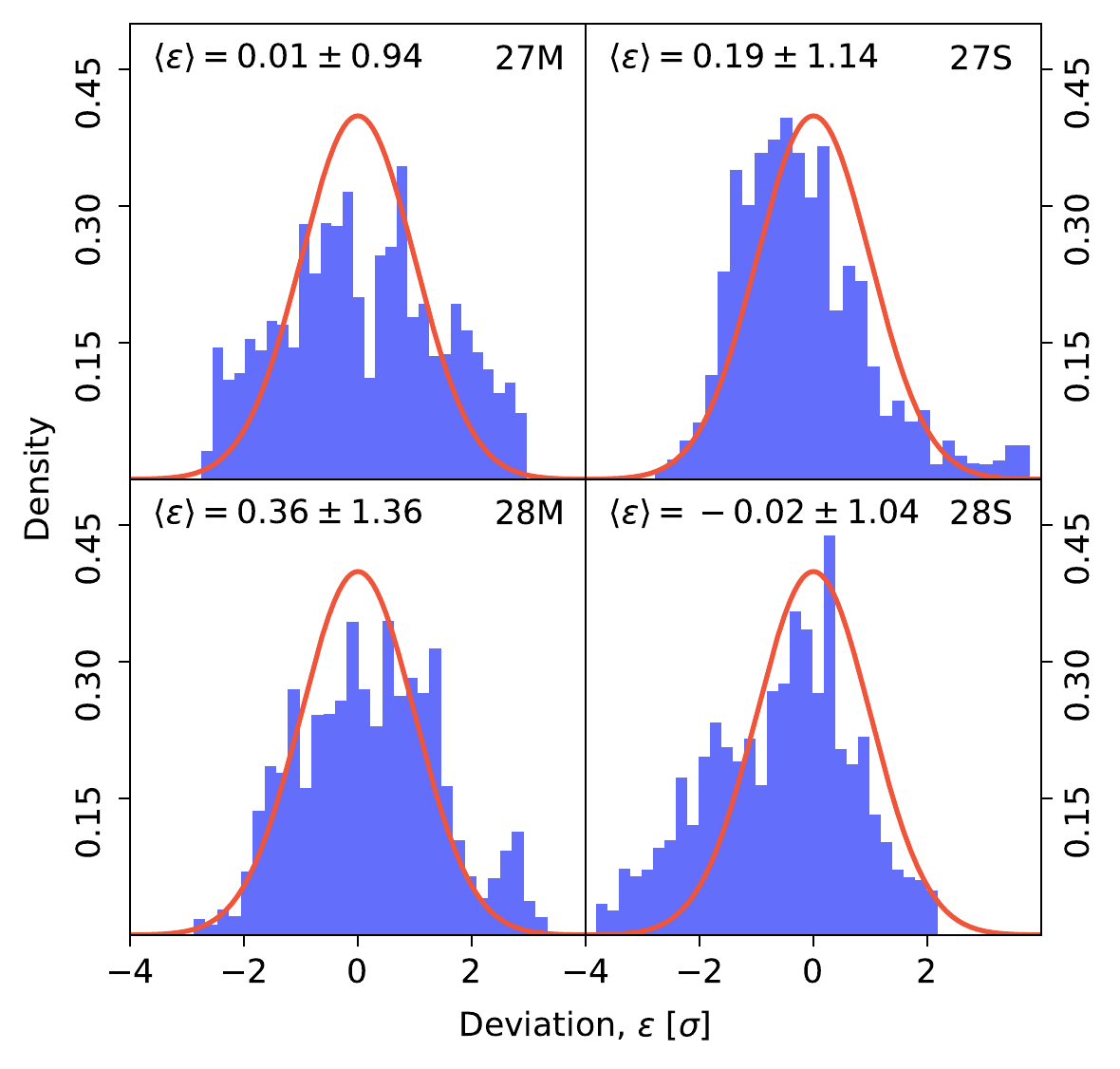}
        \caption{Aggregate standard deviation normalized differences between the gain sample mean and the input gain values. For each PID $t$ and detector $i$ we calculate $(g^{\mathrm{in}}_{t, i} - \overline{g}_{t, i}) / \sigma_{t, i}$ where $g^{\mathrm{in}}$ is the input gain, $\overline{g}$ is the mean sample value and $\sigma$ is the sample standard deviation. We then aggregate all of these values into the appropriate histogram. The red lines are ideal Gaussian distribution for comparison. Each subplot also lists the aggregate deviation from the expected mean of 0 with error bounds.}
        \label{fig:agghist_gain}

    \end{center}
\end{figure}

In contrast, the parameters in the bottom four panels have very long
correlation lengths, and these correspond to the four correlated noise
PSD parameters within a single PID; $\sigma_0$, $f_{\mathrm{knee}}$,
$\alpha$, and $A_\mathrm{p}/\sigma_0$. As discussed by \citet{bp06},
the introduction of the log-normal noise term greatly increases
degeneracies and correlations among these parameters as compared to a
standard $1/f$ noise profile, and this makes proper estimation of
these parameters much more expensive. However, it is also important to
note that this is only a challenge regarding the estimation of the individual
noise PSD parameters; the full PSD as a function of frequency,
$P_{n}(f)$, is insensitive to these degeneracies, and that function is the only thing that is actually propagated to the rest of the system. This
explains why the long correlations seen in the lower half of the plot
do not excite long correlations also among the (far more important)
parameters in the top half of the plot.

\begin{figure*}[t]
    \begin{center}
      \includegraphics[width=0.49\linewidth]{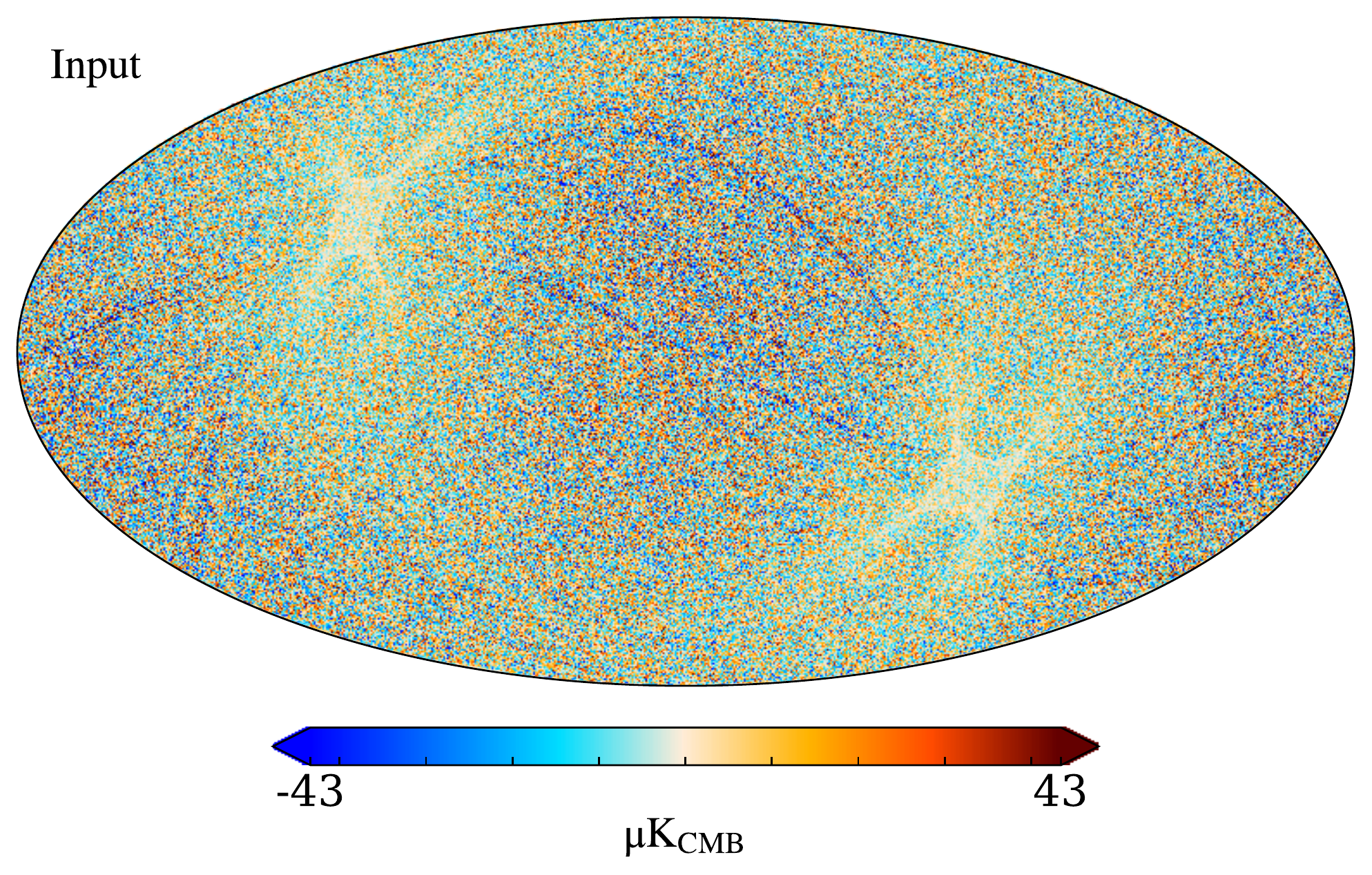}
      \includegraphics[width=0.49\linewidth]{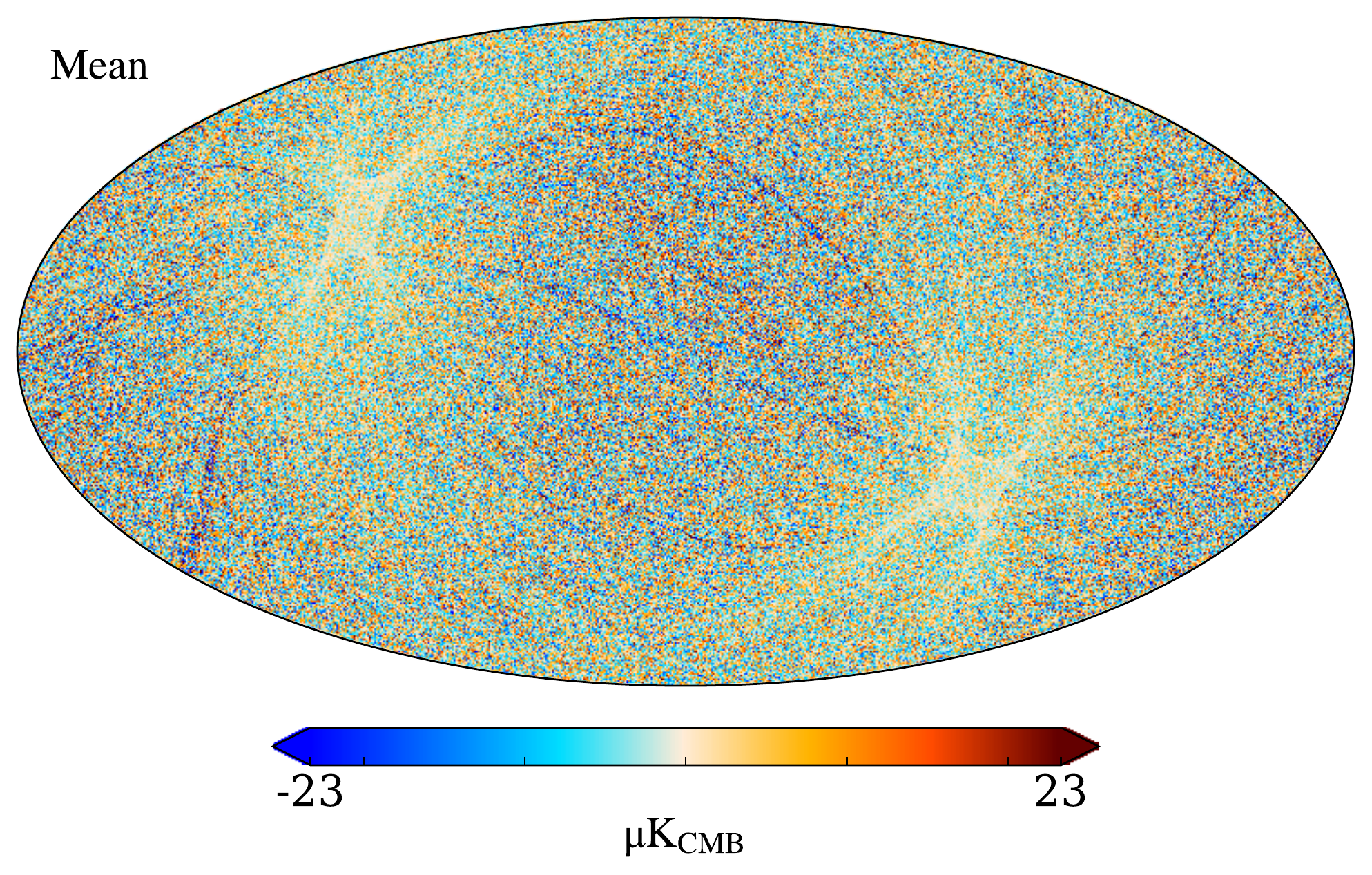}\\
      \includegraphics[width=0.49\linewidth]{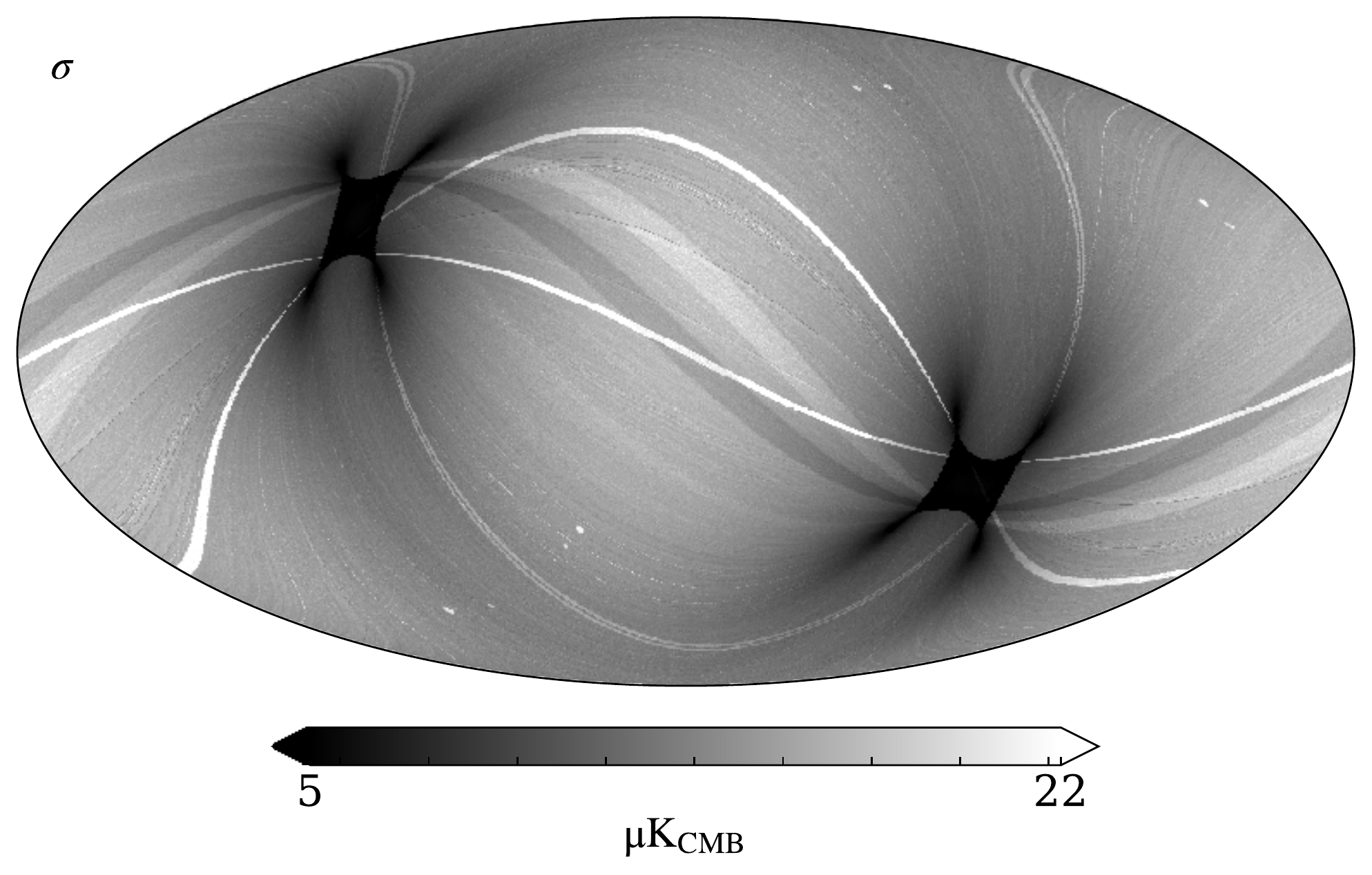}
      \includegraphics[width=0.49\linewidth]{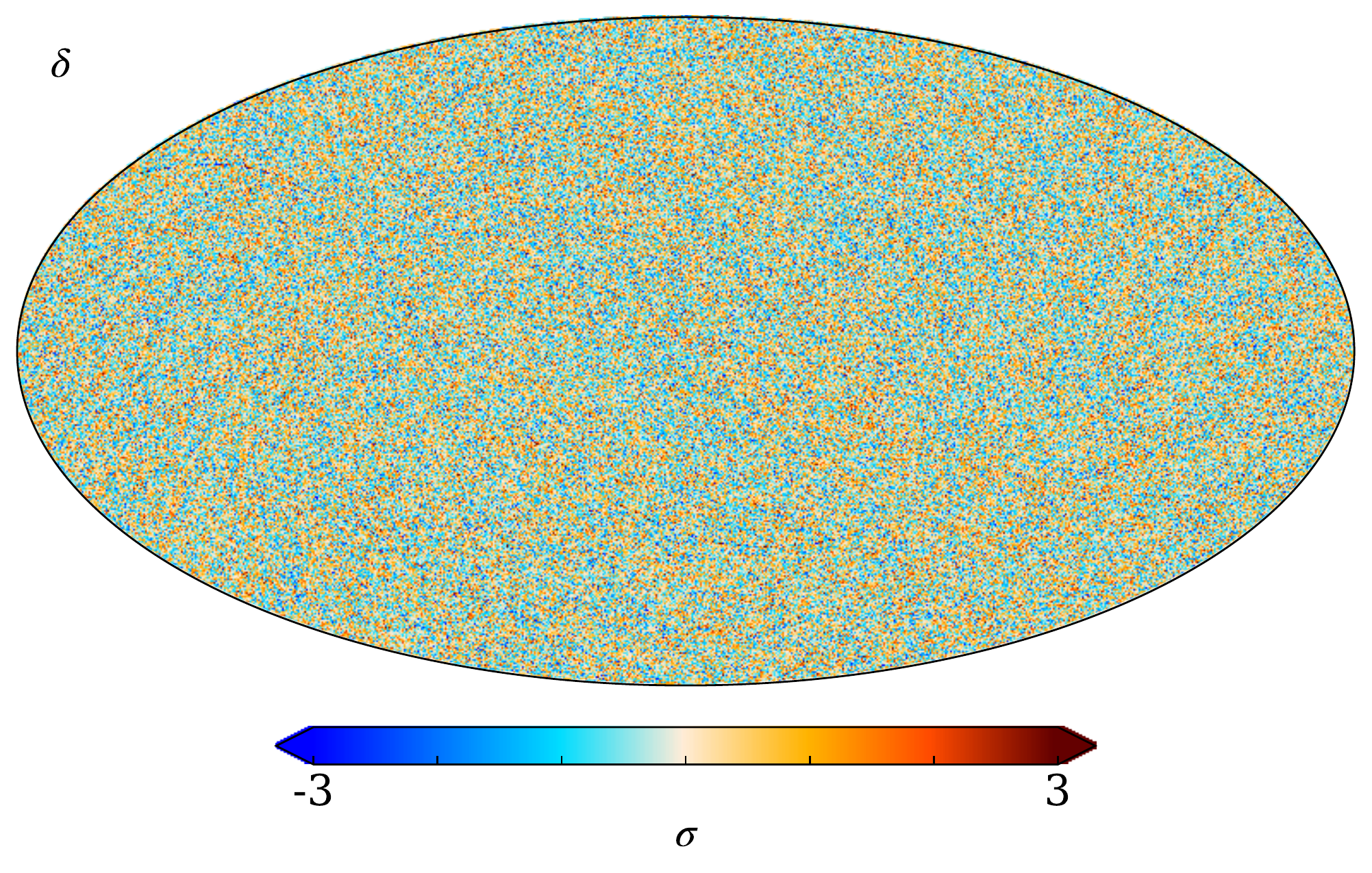}
    \end{center}
    \caption{Pixel space comparison of reconstructed correlated noise maps in temperature. (\textit{Top left}:) True input realization. (\textit{Top right}:) Estimated posterior mean (output) map.  (\textit{Bottom left}:) Estimated posterior standard deviation map. (\textit{Bottom right}:) Normalized residual in units of standard deviations.}
    \label{fig:corr_map}

    \begin{center}
        \includegraphics[width=\linewidth]{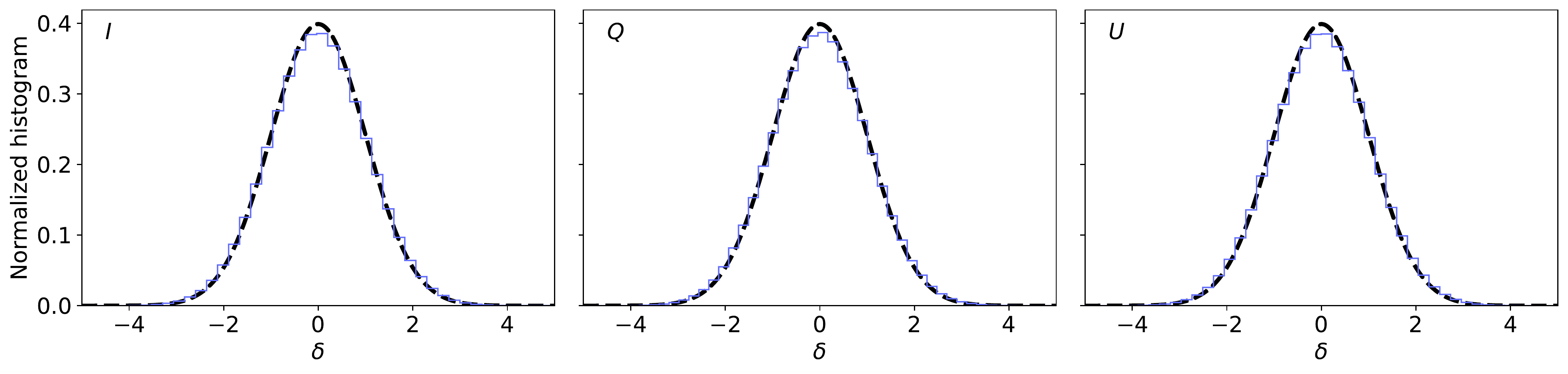}
    \end{center}
    \caption{Histograms of normalized correlated noise residuals, $\delta$ for each Stokes parameters (blue distributions).  For comparison, the dashed black line shows a standard $N(0,1)$ distribution.}
    \label{fig:corr_map_hist}
\end{figure*}

In fact, the single most important parameter in the entire system is
the CMB map, shown in the first (for individual pixels) and third (for
the quadrupole moment, $a_{2,0}$) panels. Indeed, the correlation 
length is very short or even non-existent for single pixels. This
is primarily due to the fact that this map is strongly dominated by
white noise on a single-pixel scale for the setup we consider here. As
seen in the third panel, the same does not hold true for the
quadrupole moment, in which case the correlation is in fact higher
than 0.3 at a lag of $\Delta=25$. The main driver for this is the
gain, as shown in the fourth panel. While the gain is dominated by
white noise on short time-scales (as seen by the quick drop-off
between lags of 1 and 2), there is a slow drift at higher lags. This
is caused by a partial degeneracy between the CMB map (which acts as a
calibration source in this framework, anchored by the orbital dipole)
and the overall gain. In the real \BP\ analysis, this degeneracy is
mitigated to a large extent by analyzing all LFI channels jointly, and
also by including \WMAP\ observations to break important low-$\ell$
polarization degeneracies \citep{bp07,bp10}. Still, even with those
additions there are important long-term drifts in the largest CMB
temperature scales, and these have non-negligible consequences for the
statistical significance of low-$\ell$ CMB anomalies \citep{bp11}.

\begin{figure*}[t]
  \begin{center}
    \includegraphics[width=0.49\linewidth]{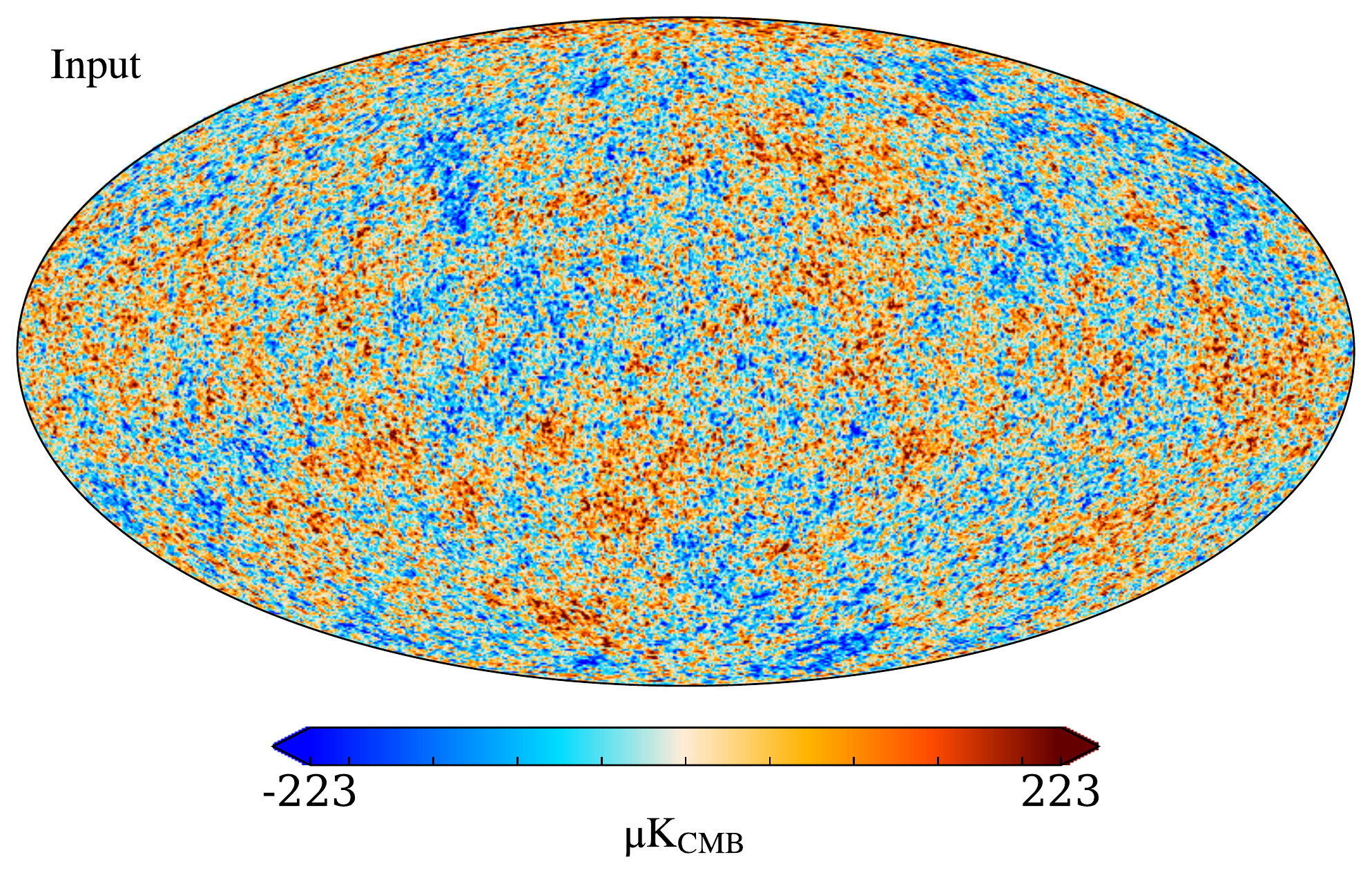}
    \includegraphics[width=0.49\linewidth]{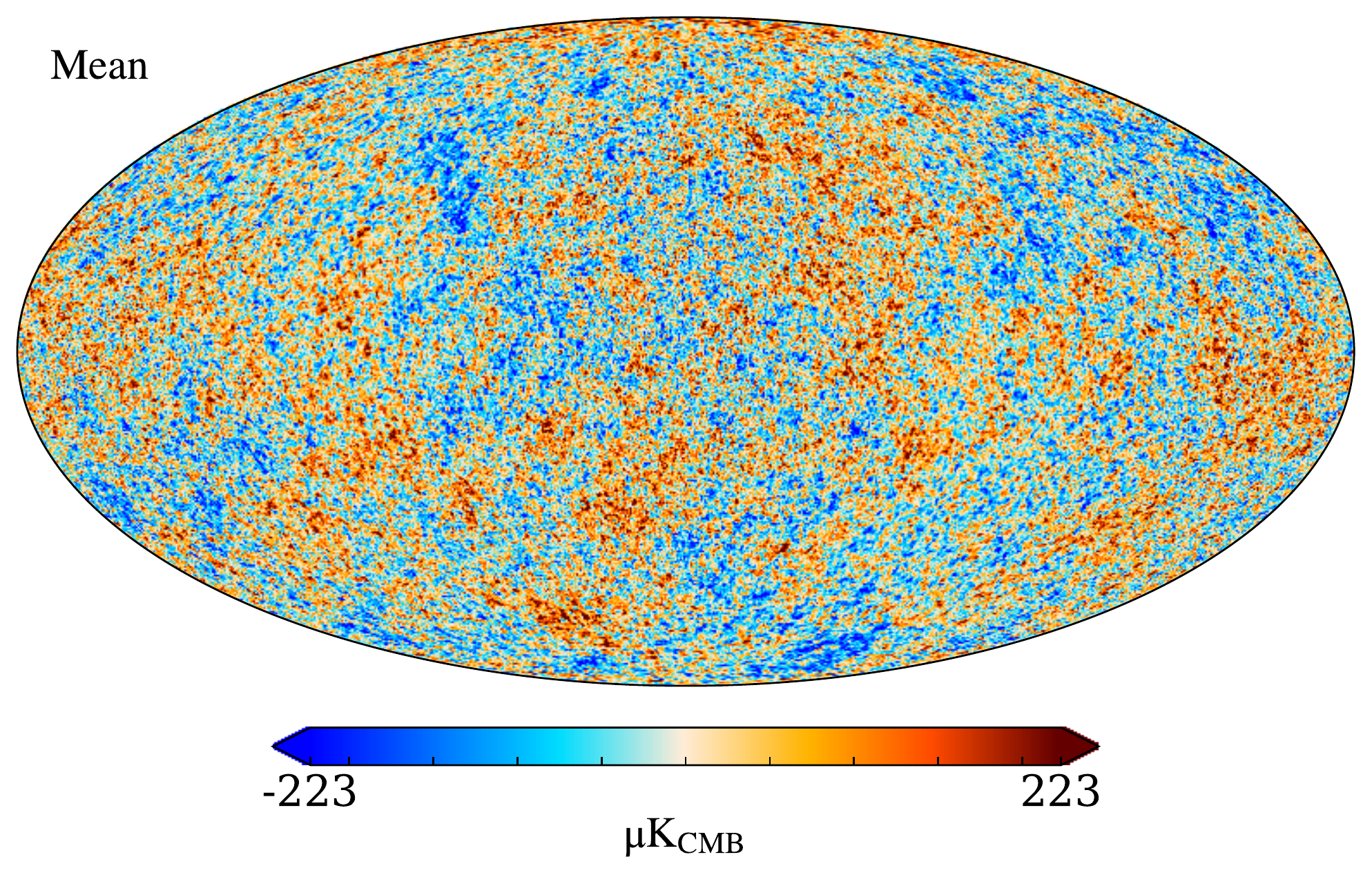}\\
    \includegraphics[width=0.49\linewidth]{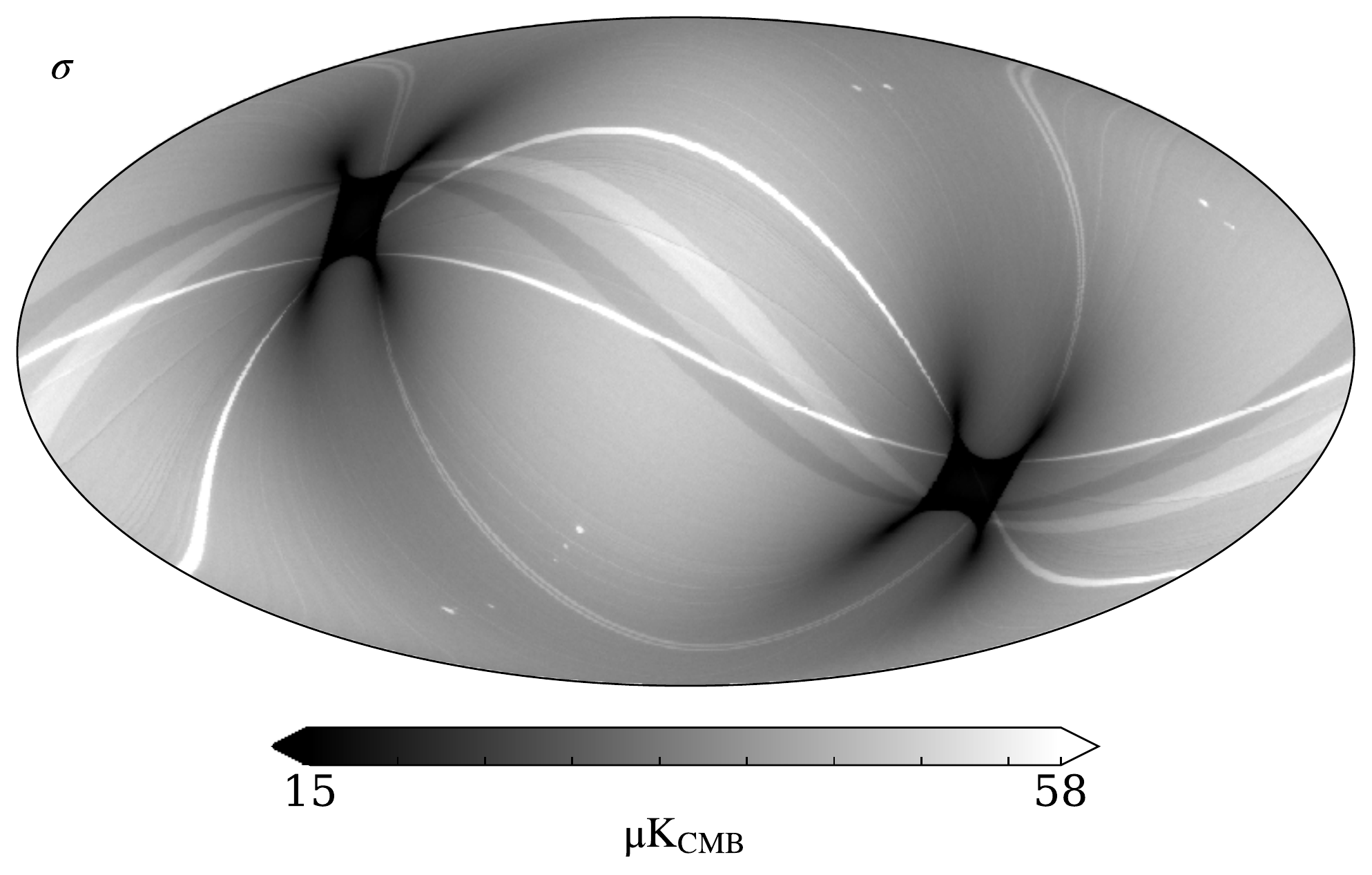}
    \includegraphics[width=0.49\linewidth]{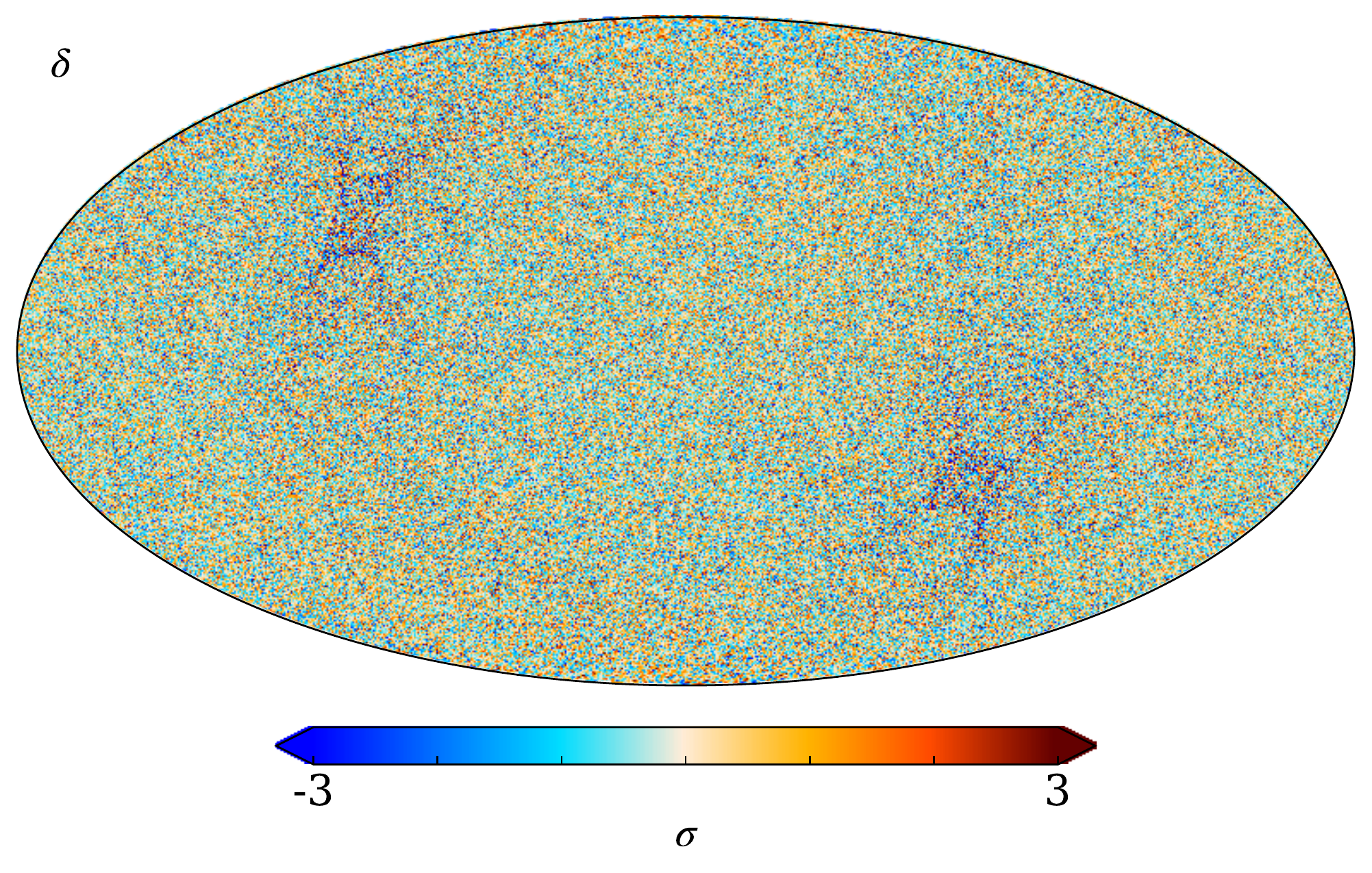}
    \end{center}
    \caption{Same as Fig.~\ref{fig:corr_map}, but for the CMB intensity component.}
    \label{fig:cmb_map}

    \begin{center}
      \includegraphics[width=\linewidth]{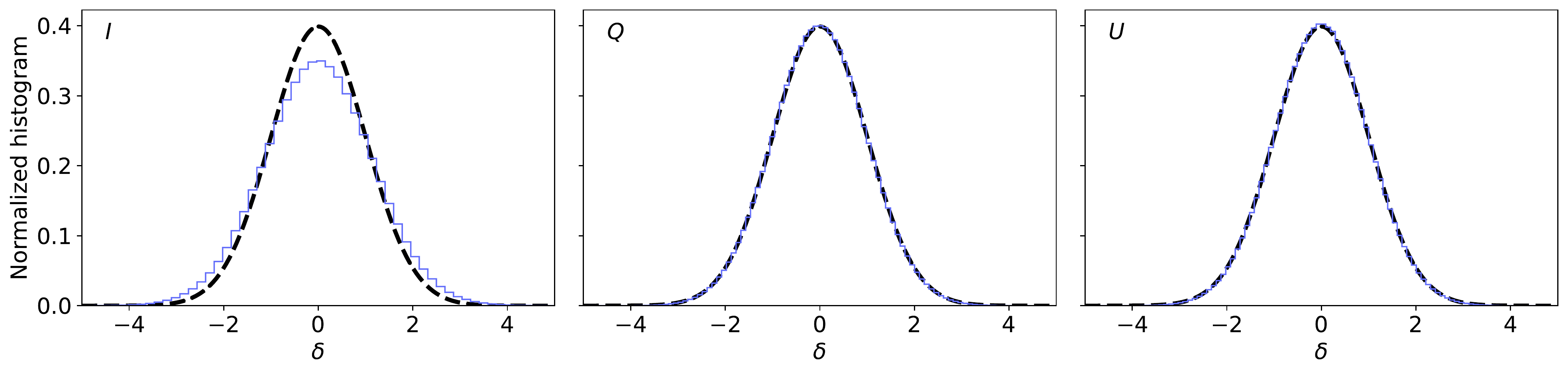}
    \end{center}
    \caption{Histograms of normalized CMB intensity residuals, $\delta_{\mathrm{CMB}}$ for each Stokes parameters (blue distributions).  For comparison, the dashed black line shows a standard $N(0,1)$ distribution.}
    \label{fig:cmb_map_hist}
\end{figure*}

\begin{figure}[t]
    \begin{center}
        \includegraphics[width=\linewidth]{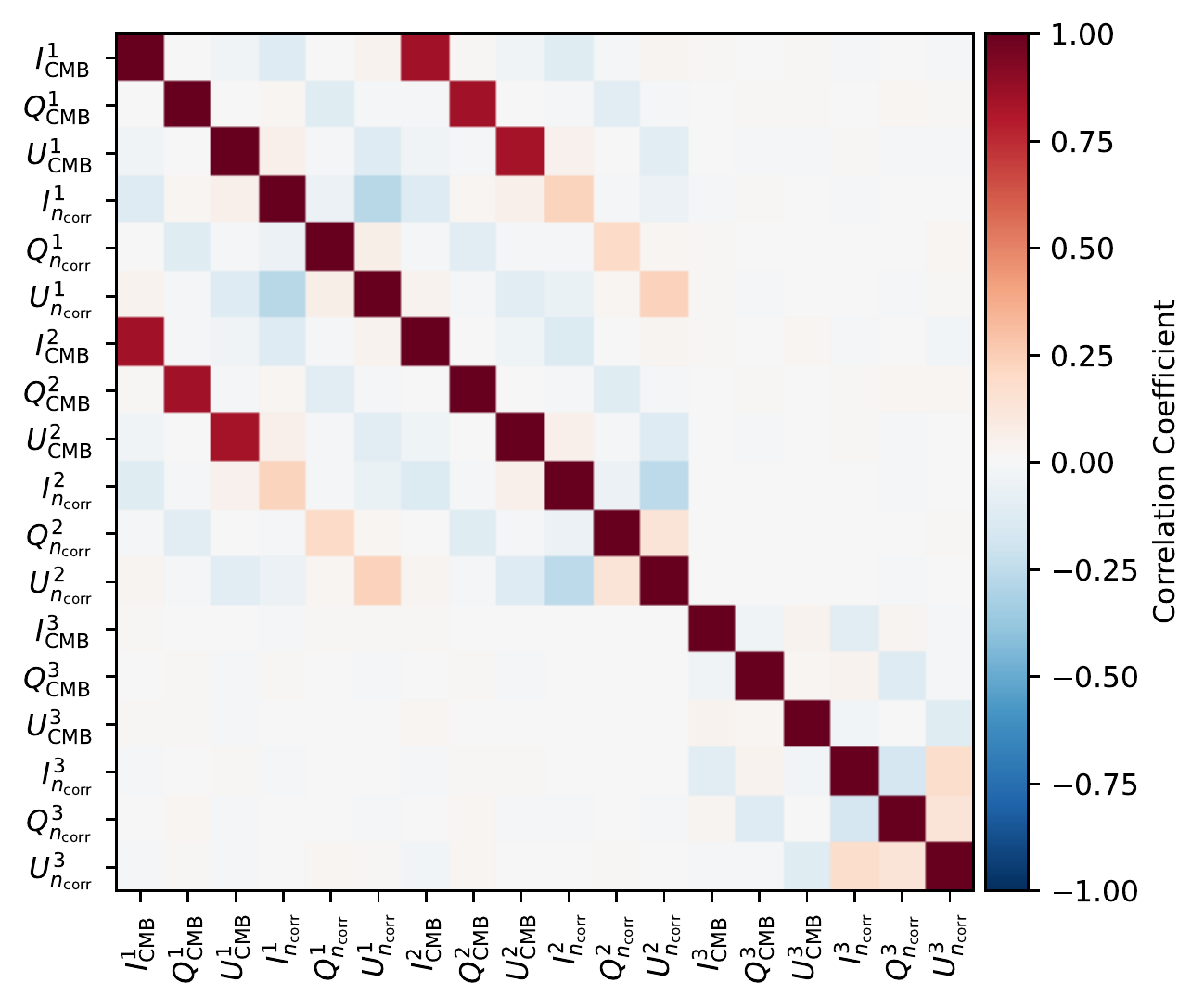}
    \end{center}
    \caption{Correlation matrix for selected pixel values of the CMB
      map, $m_\mathrm{CMB}$, and the correlated noise map,
      $m_{n_\mathrm{corr}}$, for all three Stokes parameters~$I$, $Q$, and~$U$. Pixels 1
      and 2 are selected to be neighboring pixels along the same
      \Planck\ scanning ring and located near the Ecliptic plane,
      while pixel 3 is an arbitrarily selected pixel not spatially
      associated with the other two.}
    \label{fig:pixel_corr}
\end{figure}

\subsection{Posterior distribution overview}

Next, to build intuition regarding the full set of recovered
parameters, we show in Fig.~\ref{fig:noise_corner} marginal one- and
two-dimensional posterior distributions for a small set of parameters
for two different PIDs. In each panel, the true input values are shown
as dashed lines. The bottom triangle (blue) show posterior results
for one well-behaved PID with good goodness-of-fit statistics, while
the top triangle (orange) shows a less well behaved case in
which the true input values are at the edge of recovered
distributions. Together, these two cases represent the majority of
all PIDs in terms of overall behaviour.

Overall, the true input parameters are recovered reasonably well in
most cases. One of the parameters that is less well recovered is the
white noise amplitude, $\sigma_0$. This parameter is a special case
due to the sampling algorithm currently used in the \BP\ pipeline. As
described by \citet{bp06}, $\sigma_0$ is currently determined as the
standard deviation of all pairwise differences between neighboring
time samples divided by $\sqrt{2}$. While this is a commonly used
technique in radio astronomy to derive an estimate of the white noise
that is highly robust against unmodelled systematic errors, it does
not correspond to a proper sample from the true conditional
distribution $P(\sigma_0\mid\d, \g, \ldots)$. In particular, this approach underestimates the true fluctuations of $\sigma_0$, which in turn results in the overall uncertainties being slightly underestimated.
 This is one of several examples in the pipeline in
which robustness to systematic effects comes at a cost of statistical
rigor. At the same time, it is important to note that the absolute
white noise level is in general very well determined in these data
\citep{bp06}, and a slight under-estimation of the uncertainty in
$\sigma_0$ has little practical impact on other parameters in the
model. 

Looking more broadly at the two-dimensional distributions in this
figure, we see that the parameters split naturally into two groups,
defined by the short and long correlation lengths discussed
above. That is, the CMB, correlation noise, and gain parameters
generally exhibit more symmetric distributions than the noise PSD
distributions, which are highly correlated and non-Gaussian. Once
again, this reflects the internal degeneracies among the noise PSD
parameters.

To further illustrate the impact of the slow convergence rate for
several of these parameters, Fig.~\ref{fig:triangle_chain_comparison}
shows four partial chains, each with only 1000 samples, for a sub-set
of these parameters. Once again, we see that the input values are
reasonably well recovered for most cases, but each colored
sub-distribution only cover a modest part of the full posterior
volume.

\subsection{Gain validation}

Going into greater detail on individual parameters, we show in
Fig.~\ref{fig:trace_gain} a subset of the estimated gain as a function
of Gibbs iteration for four selected PIDs, one for each
radiometer. The red lines show the true input values. Here we visually
observe the same behaviour as discussed above; on short time scales,
these trace plots are dominated by random fluctuations, while on long
time-scales there are still obvious significant drifts.

Figure~\ref{fig:band_gain} compares the estimated gain (blue bands)
with the known input (red curves) as a function of PID. The width of
the blue bands indicates the $\pm1\,\sigma$ confidence region. At
least at a visual level, the two curves agree well, without obvious
evidence of systematic biases, and the uncertainties appear
reasonable. These observations are made more quantitative in
Fig.~\ref{fig:agghist_gain}, which shows histograms of normalized
residuals, $\delta_{g}$, over all PIDs. Red lines indicate the
standard Gaussian $N(0,1)$ reference distribution. Once again, we see
that the reconstruction appears good, as the nominal bias is at most
$0.36\,\sigma$, and the maximum posterior width is
$1.36\,\sigma$. From the shape of the histograms, it is also clear
that a significant fraction of these variations is due Monte Carlo
sample variance from the long gain correlation lengths. Once again, we
note that such deviations will decrease as the number of frequency
bands included in the analysis increases, since the Solar CMB dipole,
which is the main calibrator, will be much better constrained with
more observations; the actual gain correlation lengths found for the
real \BP\ analysis are shown by \citet{bp07}, and are notably shorter
than those of this reduced simulation.

\subsection{Correlated noise posterior validation}
\label{sec:ncorr}

Next we turn to the correlated noise component, and we start with the
specific noise realization, $\n_{\mathrm{corr}}$; the correlated noise
PSD parameters will be discussed separately in
Sect.~\ref{sec:corr_psd}. To simplify the visualization, we bin the
correlated noise TOD into a sky map, as illustrated in
Fig.~\ref{fig:corr_map}. The top left panel shows the true input
correlated noise map (temperature component only), while the top right
panel shows the corresponding posterior mean (output) map. The bottom
left panel shows the posterior standard deviation per pixel, and the
bottom right panel shows the normalized residual,
$\delta_{\mathrm{corr}}$.

A visual inspection of the simulation input and posterior mean
correlated noise maps indicates no obvious differences. In fact, the
normalized residual map in the bottom right panel of 
Fig.~\ref{fig:corr_map} appears fully
consistent with white noise. Once again, this observation is
quantified more accurately in Fig.~\ref{fig:corr_map_hist}, where we
compare the histogram of $\delta_{\mathrm{corr}}$ over all pixels with
the usual $N(0,1)$ distribution for each of the three Stokes
parameters; in each case, the agreement is excellent.

\subsection{CMB map validation}

Figures~\ref{fig:cmb_map} and \ref{fig:cmb_map_hist} show similar
plots for the CMB sky map component. Once again, the normalized
residual in the bottom right panel appears fully consistent with white
noise over most of the sky --- but this time, we actually see a power
excess in $\delta_{\mathrm{CMB}}$ around the Ecliptic poles. These
features correspond to regions of the sky that are particularly deeply
observed by the \Planck\ scanning strategy \citep{planck2013-p01}. As
a result of these deep measurements, the white noise in these regions
is very low, and the total error budget per pixel is far more
sensitive to the non-linear contributions in the system, in particular
the coupling between the gain and the Solar dipole.

This effect does of course not only apply to the Ecliptic ``deep
fields'', but to all signal-dominated map pixels at some level, and it
therefore also applies to the full-sky CMB map in temperature. This
statement is made more quantitative in the left panel of
Fig.~\ref{fig:cmb_map_hist}, where we see that the temperature
histogram is very slightly wider than the reference $N(0,1)$
distribution. To be specific, the standard deviation of this
distribution is about 1.15; at the same time, the mean of the
distribution is consistent with zero, the non-linear couplings
therefore do not introduce a bias, but only a higher variance. For the
noise-dominated Stokes $Q$ and $U$ parameters, for which gain
couplings are negligible on a per-pixel level, both distributions are
perfectly consistent with $N(0,1)$.

Figure~\ref{fig:pixel_corr} shows Pearson's correlation coefficients
between the CMB and correlated noise components for three selected
pixels. Two of the pixels, marked `1' and `2', are located along the
same \Planck\ scanning ring near the Ecliptic plane, where the
\Planck\ scanning strategy is particularly poor. The third pixel is
located far away from these, and on a different scanning ring. Here we
see that correlations are very strong for Stokes parameters of the
same type along the same ring, with correlation coefficients ranging
between 0.5 and 0.8. These correlations are induced both by gain and
correlated noise fluctuations, which are tightly associated with the
\Planck\ scanning rings. Stokes parameters of different types (e.g. I and Q) are significantly less correlated, typically with anti-correlation
coefficients of $\rho\lesssim-0.25$. Correlations between widely
separated pixels are practically negligible in the current simulation
setup, although for the real analysis, this is no longer true due to
additional couplings from, for instance, astrophysical foregrounds,
bandpass corrections, and sidelobes \citep{bp08,bp09,bp10,bp11,bp13,bp14}.

\begin{figure}[t]
    \includegraphics[width=\linewidth]{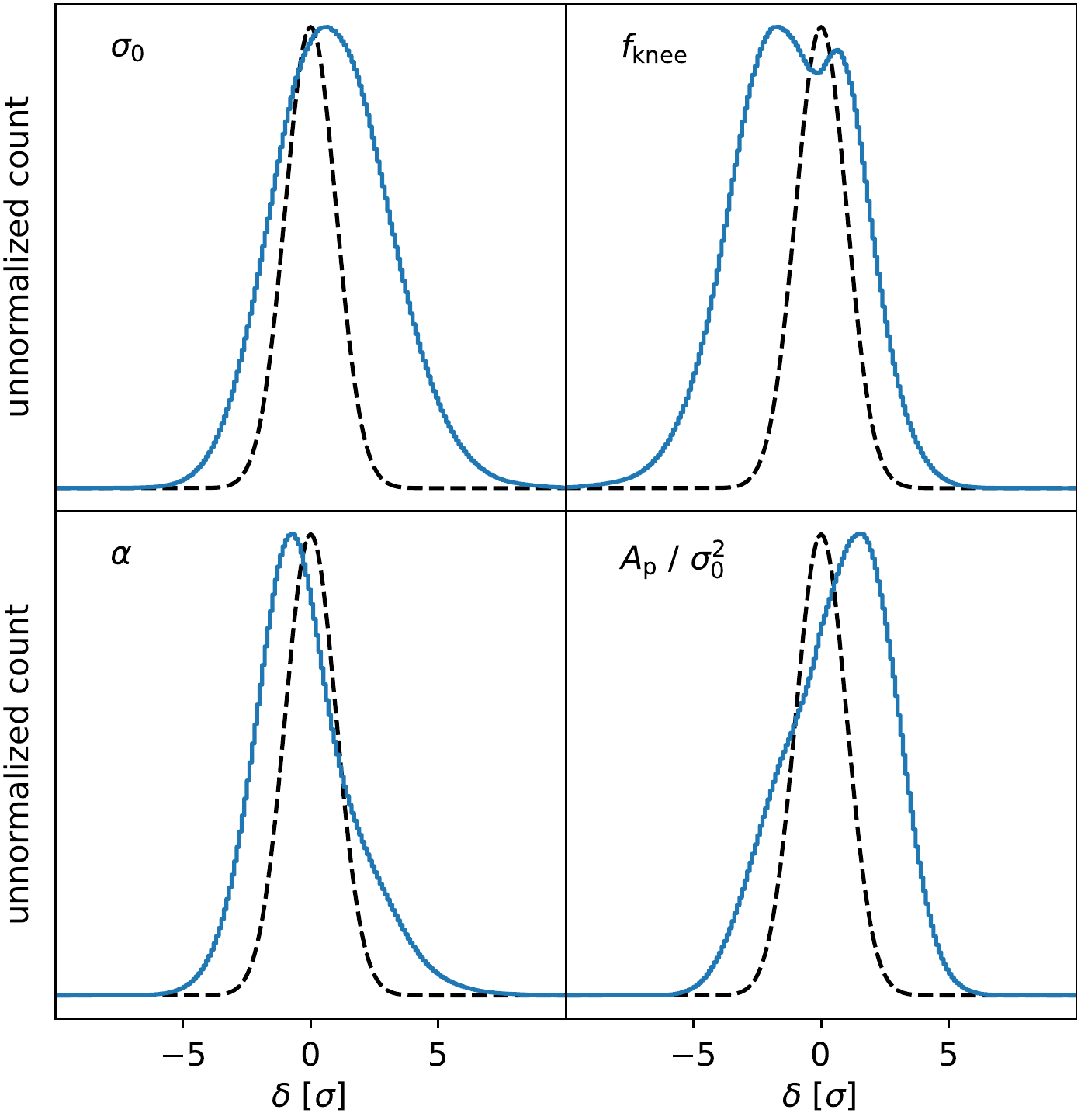}\\
    \caption{Histograms of the noise parameters over all PIDs and 10\,000 samples for radiometer 27M. We show the white noise level~$\sigma_0$, knee
      frequency~$f_\mathrm{knee}$, correlated noise spectral
      index~$\alpha$, and log-normal noise
      amplitude~$A_\mathrm{p}$. For reference we show the standard normal distribution as a black dashed line.}
    \label{fig:xi_n}
\end{figure}


\begin{figure}[t]
  \center
  \includegraphics[width = \linewidth]{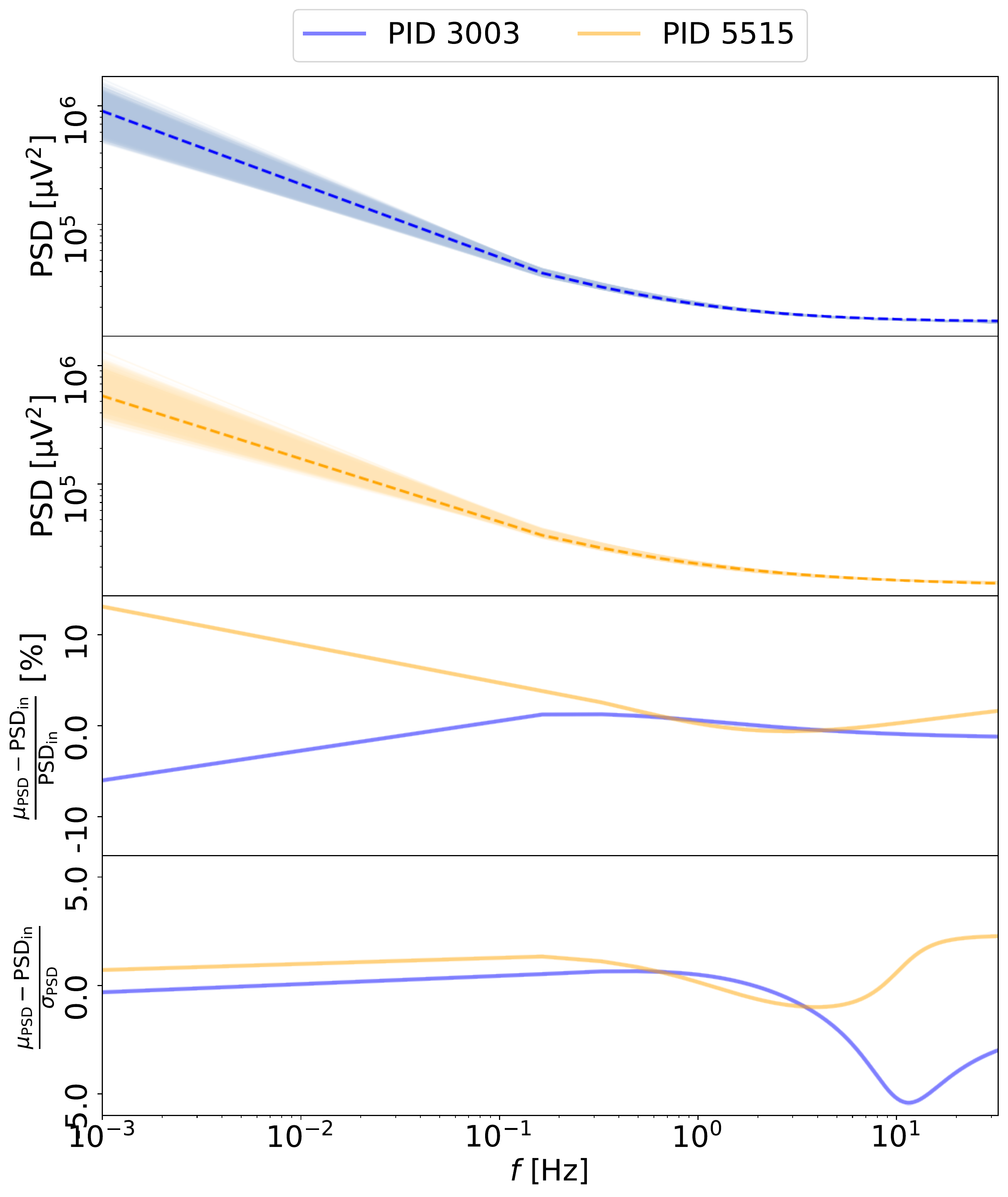}
  \caption{Comparison of recovered correlated noise PSD in terms of
    the functional form, $P_n(f)$. The top two panels show results for
  the same PIDs as in Fig.~\ref{fig:noise_corner}; faint lines
  indicate individual Gibbs samples, while the dashed lines show the
  true input functions. The bottom two panels show the difference
  between the posterior mean function and the true input in units of
  percent and posterior rms, respectively. }
  \label{fig:noise_psd}
\end{figure}

\subsection{Correlated noise PSD validation}
\label{sec:corr_psd}

Finally, we consider the noise PSD parameters, $\sigma_0$,
$f_\mathrm{knee}$, $\alpha$, and $A_\mathrm{p}/\sigma_0$, and, as
already noted, these are significantly harder to estimate individually
than the previous parameters due to the strong correlation between the
$1/f$ and log-normal terms in Eq.~\eqref{eq:1fmodel_lognorm}.

As usual, we plot the reduced residual, $\delta$, for each parameter
type in Fig.~\ref{fig:xi_n}, and in this case we see that the
posterior distributions are significantly wider than standard Gaussian
distribution, by as much as a factor of two. The distributions are
also clearly non-Gaussian, with notable skewness and
kurtosis. Both the excess variance and non-Gaussianity stem from the
same degeneracies as discussed above, and are partially due to
intrinsic non-Gaussianities in the model, and partially due to
incomplete Monte Carlo convergence and very long correlation
lengths. On the other hand, the mean bias in these distribution is
small, and the estimated posterior distributions do provide a useful
summary of each parameter individually.

As mentioned above, however, other parameters in the model are not
sensitive to individual $\xi^n$ values, but only to the total
noise PSD, $P_\mathrm{corr}(f)$. This function is plotted in
the top two panels of Fig.~\ref{fig:noise_psd} for the same two PIDs and radiometers as shown in
Fig.~\ref{fig:noise_corner}; the blue curves correspond to the
well-measured PID, while the orange curve corresponds to the PID with
the marginal fit. Faint lines in the top two panels show individual
Gibbs samples, corresponding to different combinations of $\xi^n$. By
eye, the sampled values appear to span the true input reasonably well,
although the orange line is on the lower edge of the estimated
posterior distribution.

These visual observations are made more quantitative in the bottom two
panels, where the third panel shows the fractional difference between
the output and input PSD functions in units of percent, and the fourth
panel shows the same in units of standard deviation of the PSD across 
Gibbs samples, $\sigma$. For the well-behaved (blue)
pixel, we see that the posterior mean matches the true input
everywhere to within a few percent; in units of standard deviations,
this is typically less than $2.5\,\sigma$ for most of the region,
except at frequencies above 10\,Hz, where the estimated standard
deviation is very small due, and the underestimation of the
uncertainty in $\sigma_0$ becomes noticeable. For the less
well-behaved case, the recovered PSD is within $2\,\sigma$ at all
frequencies in units of standard deviations, or within 5\,\%. Overall,
the PSD itself is recovered very well in both cases in absolute terms.

\section{Conclusions}
\label{sec:conclusions}

End-to-end time-ordered simulations play a key role in estimating both
biases and uncertainties for current and future CMB experiments. To date, no other practical method has been able to
account for the full and rich set of systematic errors that affect
modern high-precision measurements.

As detailed by \citet{bp01} and its companion papers, the \BP\ project
has implemented a new approach to end-to-end CMB analysis in which a
global parametric model is fitted directly to the time-ordered data,
allowing for joint estimation of instrumental, astrophysical, and
cosmological parameters with true end-to-end error propagation. This
approach relies strongly on a sampling algorithm called Gibbs
sampling, which allows the user to draw joint samples from a complex
posterior distribution. Each of these Gibbs samples correspond
essentially to one end-to-end TOD simulation, similar to those
produced by classical CMB simulation pipelines, for instance the
\Planck\ full focalplane (FFP; \citealp{planck2014-a14})
simulations.

The fundamental difference between these two simulation pipelines lies
in how to define the input parameters used to generate the
simulation. In the \BP\ approach, all parameters are constrained
directly from the true data, and correspond as such to samples drawn
from the full joint posterior distribution. In contrast, traditional
pipelines uses parameters that are a mixture of data-constrained and
data-independent parameters. Typical examples of the former include
the CMB Solar dipole and Galactic foregrounds, while typical examples
of the latter include CMB anisotropies and instrumental noise. In this
paper, we have the two types of simulations for ``Bayesian'' or
``frequentist'', respectively, indicating whether or not they
condition on the true data.

The difference between these two types of simulations has direct
real-world consequences for what applications each simulation type is
suitable for. As first argued by \citet{bp10}, this may be intuitively
understood through the following line of reasoning: Suppose one is
tasked with constructing a new end-to-end simulation for a given
experiment. Among the first decisions that needs to taken concerns the
CMB Solar dipole: Should this correspond to the true dipole, or should
it have a random amplitude and direction? If it is chosen randomly,
then the hot and cold spots in the correlation matrices shown in
Fig.~\ref{fig:ncov} in this paper will appear at random positions on
the sky, and eventually be washed out in an ensemble average. In
practice, all current pipelines adopt the true CMB Solar dipole as an
input. The next question is, what Galactic model should be used? Once
again, if this is selected randomly, then the Galactic plane will move
around on the sky from realization to realization. In practice, all
current pipelines adopt a model of the true Galactic signal as an
input.

The third question is, what CMB anisotropies should be used? At this
point, all pipelines prior to \BP\ have in fact adopted random CMB
skies drawn from a theoretical $\Lambda$CDM model. This has two main
effects: On the one hand, in the same way that randomizing the CMB
dipole signal would average out any coherent correlations between the
sky signal and the gain, randomizing the CMB anisotropies also average
out, and non-linear correlations between these structures and the
instrumental parameters are not accounted for. On the other hand, the
resulting simulations do actually include so-called cosmic variance,
i.e., for the scatter between individual CMB realizations.

Fourth and finally, the same question apply to all the instrumental
parameters, perhaps most notably correlated noise and gain
fluctuations: Should these be constrained by the real data, or should
they be drawn randomly from a laboratory-determined
hyper-distribution? 

It is important to stress that none of these four questions have a
``right'' or ``wrong'' answer. However, whatever choice one makes,
that choice will have direct consequences for what correlation
structures appear among the resulting simulations, and therefore also
for which applications they are suitable for. In particular, if the
primary application is traditional frequentist model testing --- for
instance asking whether the CMB sky is Gaussian and isotropic --- then
it is critical to account for cosmic variance among the CMB
realizations. For those applications, one must choose data-independent
CMB inputs in order to capture the full uncertainties, and the
appropriate choice are frequentist data-independent simulation inputs.

If, on the other hand, the main application is traditional parameter
estimation, for instance as constraining the $\Lambda$CDM model, then
the important point is to properly estimate the total CMB uncertainty per-pixel on the sky. In this case, it is
critical to properly model all non-linear couplings between the actual
sky signal, the true gain, the true correlated noise, etc. In this
case, the appropriate choice are Bayesian data-dependent simulation
inputs.

In this paper, we note that the novel \commanderthree\ software is
able to produce both frequentist and Bayesian simulations, simply by
adjusting the inputs that are used to initialize the code. While the
Bayesian simulation process has been described in detail in most of
the \BP\ companion papers, we present in the current paper a first
application of the frequentist mode of operation by producing a
data-independent time-ordered simulation corresponding to one year of
30\,GHz data, and we then use this to validate three important
low-level steps in the full \BP\ Gibbs samples, namely gain
estimation, correlated noise estimation, and mapmaking. Doing so, we
find that the recovered posterior distribution matches the true input
parameters well.

\begin{acknowledgements}
  We thank Prof.\ Pedro Ferreira and Dr.\ Charles Lawrence for very
  useful suggestions, comments and discussions.  We also thank the
  entire \Planck\ and \WMAP\ teams for invaluable support and
  discussions, and for their dedicated efforts through several decades
  without which this work would not be possible. This research has
  partially been carried out within the master- and PhD-level course
  ``AST9240 --- Cosmological component separation'' at the University
  of Oslo, and support for this has been provided by the Research
  Council of Norway through grant agreement no.\ 274990. The
  BeyondPlanck Collaboration has received funding from the European
  Union’s Horizon 2020 research and innovation programme under grant
  agreement numbers 776282, 772253, and 819478. In addition, the
  collaboration acknowledges support from ESA; ASI, CNR, and INAF
  (Italy); NASA and DoE (USA); Tekes, AoF, and CSC (Finland); RCN
  (Norway); ERC and PRACE (EU).  L.\,T.\,Hergt was supported by a UBC
  Killam Postdoctoral Research Fellowship.  The work of Tamaki
  Murokoshi was supported by MEXT KAKENHI Grant Number 18H05539 and
  the Graduate Program on Physics for the Universe (GP-PU), Tohoku
  University.  Part of the works by F.\, Rahman was carried out using
  the \texttt{Nova} and \texttt{Coolstar} clusters at the Indian
  Institute of Astrophysics, Bangalore.  The work of
  K. S. F. Fornazier, G. A. Hoerning, A. Marins and F. B. Abdalla was
  developed in the brazilian cluster SDumont and was supported by
  S\~ao Paulo Research Foundation (FAPESP) grant 2014/07885-0,
  National Council for Scientific and Technological (CNPQ),
  Coordination for the Improvement of Higher Education Personnel
  (CAPES) and University of S\~ao Paulo (USP).  Some of the results in
  this paper have been derived using the \texttt{healpy}
  \citep{zonca2019} and \texttt{HEALPix} \citep{gorski2005} packages.
  This work made use of Astropy\footnote{http://www.astropy.org}
  \citep{astropy:2013, astropy:2018, astropy:2022}.
\end{acknowledgements}

\bibliographystyle{aa}

\bibliography{common/Planck_bib,common/BP_bibliography}

\end{document}